\documentclass[traditabstract]{aa}

\usepackage{graphicx}
\usepackage{txfonts}

\usepackage{epsfig}
\usepackage{natbib}

\usepackage{aalongtable}

\begin{document}

\title{The EUV spectrum of the Sun: quiet and active Sun irradiances
and chemical composition}
\author{G. Del Zanna 
}


\institute{DAMTP, Centre for Mathematical Sciences,  
 University of Cambridge,  Wilberforce Road, Cambridge CB3 0WA UK
}

 \date{ Accepted for publication}

 \abstract{We benchmark new atomic data against a selection of irradiances 
obtained from medium-resolution 
quiet Sun spectra  in the EUV, from 60 to 1040~\AA. 
We use as a baseline the irradiances measured during solar 
minimum on 2008 April 14 by the prototype (PEVE) of the 
Solar Dynamics Observatory Extreme ultraviolet Variability Experiment (EVE).
We  take into account some inconsistencies in the PEVE data, using 
flight EVE data  and irradiances we obtained from
Solar \& Heliospheric Observatory (SoHO) 
 Coronal Diagnostics Spectrometer (CDS) data. We perform a 
differential emission measure and find overall excellent agreement 
(to within the accuracy of the observations, about 20\%)
between predicted and measured irradiances in most cases,
although we point out several problems with the currently available 
ion charge state distributions.
We used the  photospheric chemical abundances of Asplund et al. (2009).
The new atomic data are nearly complete in this spectral range,
for medium-resolution irradiance spectra.   
Finally, we use  observations of the active Sun in 1969 to show that 
also in that case the composition of the solar corona up to 
1 MK is nearly photospheric. Variations of a factor of 2 
are present for higher-temperature plasma, which is emitted within 
active regions. These results are in excellent agreement with our previous 
findings.
 \keywords{Sun: abundances -- Sun: corona -- Sun: UV radiation -- Line: identification -- Techniques: spectroscopic }
 }

\maketitle

\section{Introduction }

This paper  is one of a series of studies  where 
the EUV radiances and irradiances of the Sun are studied. 
One of the goals of these studies 
was the provision of well-calibrated  EUV/UV  irradiances of the quiet and active Sun,
to aid the interpretation and modelling of stellar observations. 
 As a first step on the modelling 
side, it is therefore important to  assess   the completeness and accuracy of  current
atomic data. This is the main aim of the present paper, where
 a large amount of recently-calculated atomic rates are included.
We focus  on   quiet Sun  irradiances, but also 
point out the differences with those of the more active Sun.

The present benchmark also provides a test on ion charge state distributions.
In fact, to model full-Sun irradiances, it is well known that a 
continuous emission measure distribution is obtained. Therefore, lines
of different ions formed at similar temperatures should have 
predicted irradiances close to the observed ones. Any discrepancies 
turn out to be mostly due to problems in the ion charge state calculations.
Finally, the present benchmark also provides an excellent 
way to measure the chemical abundance of the solar transition region and corona.

An earlier assessment  was carried out within the 
SOLID\footnote{http://projects.pmodwrc.ch/solid/} project,
which aimed at providing and estimating the solar spectral 
irradiance across all wavelengths.
Within SOLID, a comparative study between the EUV/UV spectral 
radiances/irradiances as observed by
several  instruments and as obtained with modelling 
using  CHIANTI\footnote{www.chiantidatabase.org} atomic data \citep{dere_etal:97} was carried out. 
The earlier assessment was used to establish  which atomic data needed improvement
and which were missing, and ultimately led to a significant update of the 
CHIANTI database, version 8 \citep{delzanna_etal:2015_chianti_v8}.

As most of the issues discussed in the previous series of papers is 
relevant to the present analysis, we provide a brief summary:
Paper I  \citep{delzanna_etal:2010_cdscal}  
presented  a new radiometric calibration of the 
 Solar \& Heliospheric Observatory (SoHO)  Coronal Diagnostics Spectrometer  
\citep[CDS;][]{Harrison-etal:95} Normal Incidence Spectrometer (NIS), 
which included earlier calibrations \citep{delzanna_etal:2001_cdscal,brekke_etal:00}.
In Paper I, quiet Sun NIS irradiances in 2008  were obtained and compared to 
 those measured by  a sounding rocket,  flown on  2008 April 14 during the extended 
solar minimum, when the Sun was very quiet. 
The rocket carried a prototype (PEVE, see \citealt{woods_etal:09})  of the
the Solar Dynamics Observatory (SDO) 
 Extreme ultraviolet Variability Experiment (EVE, see \citealt{woods_etal:12}),
and produced a medium-resolution EUV spectrum 
over a broad  spectral range (60--1040~\AA).
Very good agreement between the NIS and PEVE irradiances for most of the 
lines was found.
However,  problems in the PEVE irradiances  in some of the strongest lines were found.
This was surprising as the PEVE instrument was radiometrically calibrated 
on the ground \citep{chamberlin_etal_07,hock_etal:2010,chamberlin_etal:09}.

Our new SoHO CDS calibration allowed the first measurements of the 
EUV spectral irradiance along a solar cycle, from 1998 until 2010, which was 
 presented in Paper II \citep{delzanna_andretta:2011}.
The irradiances in lines formed below 1 MK were shown to vary little, with the
exception of the He lines. 
A new calibration for the  most prominent line  in the EUV,
 the  \ion{He}{ii} 304~\AA\ resonance transition, 
was proposed. The change was significant, and implied that 
most previous calibrated values were incorrect by about a factor of two.
Our new  NIS calibration was later confirmed by a 
sounding rocket flight, EUNIS (see \citealt{wang_etal:2011}).
Paper III \citep{andretta_delzanna:2014} analysed in detail the radiances
and limb-brightening in the main lines and their variation with the solar cycle. 
Paper IV \citep{delzanna_etal:2015_SEM} compared the CDS irradiances with the
count rates measured
by the Solar EUV Monitor (SEM)  first-order band (SEM~1).
Paper V  \citep{delzanna_andretta:2015}
extended and revised the CDS NIS calibration  until 2014
when the instrument was switched off.
The irradiances obtained from this analysis were then compared to those
measured by  PEVE, by  SDO EVE and by the
Thermosphere Ionosphere Mesosphere Energetics Dynamics
(TIMED)  Solar EUV Experiment (SEE) EUV Grating Spectrograph (EGS)
 \citep{woods_etal:05}.
 Excellent agreement (to within  a relative 10--20\%) 
with the  EVE data (only during the 2010--2012 period) was found;
however, the  problems in  the PEVE irradiances were confirmed, and
several discrepancies with the  TIMED EGS  irradiances were also found.

The present assessment is the first of its kind in the EUV as briefly 
pointed out  below in Section~2. Section~3 briefly describes the new atomic data used in the 
present paper, while Section~4 presents the quiet Sun irradiances, a 
differential emission measure (DEM) modelling
and predicted spectra.
Section~5 revisits older irradiance observations of the active Sun,
while Section~6 draws the conclusions.

\section{Previous benchmark studies in the EUV}

 Several assessments of atomic data in the EUV 
 have been carried out previously, but on radiances
in narrow wavelength ranges or in specific ions. For example, 
in a long series of papers starting with \cite{delzanna_etal:04_fe_10}, 
the atomic data of several key ions have been benchmarked against 
laboratory and astrophysical spectra, providing many lacking identifications in 
the EUV, as reviewed in  \cite{delzanna_mason:2018}. 
Most of these identifications have been included in  CHIANTI
version 7.1 and 8. 

A significant benchmark study  based on the first 
SERTS rocket flight \citep[][]{thomas_neupert:94} was carried out by 
\cite{young_etal:98} in the 170-450~\AA\ range.
The most complete previous assessment of atomic data for the EUV used 
SoHO CDS NIS and grazing incidence (GIS) radiances  in the 150--780~\AA\ range
 \citep{delzanna_thesis:1999}, but was limited by an uncertain 
CDS radiometric calibration and the use of older CHIANTI atomic data.

\cite{landi_etal:2002_cds} also performed a benchmark of CHIANTI data 
for the coronal lines using off-limb observations from  CDS NIS.
An assessment of more recent CHIANTI atomic data in the 166--212~\AA\ and  245--291~\AA\
spectral ranges observed by the  Extreme-ultraviolet Imaging Spectrometer 
(EIS, see \citealt{culhane_eis:07}) aboard Hinode
was carried out in  \cite{delzanna:12_atlas} on coronal lines.
\cite{delzanna:09_fe_7} and \cite{landi_young:2009}  benchmarked CHIANTI atomic data 
 for transition-region lines observed by EIS.

SoHO Solar Ultraviolet Measurements of Emitted Radiation
(SUMER, \citealt{Wilhelm-etal:95}, 700--1600~\AA)  
quiet Sun radiances  were compared with earlier CHIANTI atomic data 
by \cite{doschek_etal:1999}, finding significant discrepancies.
We note, however, that the lack of simultaneity in observing SUMER lines 
far apart in wavelength is a limiting issue for this type of studies,
considering that the dominant lines are highly variable. 
An improvement was the benchmark of CHIANTI data for coronal lines 
using quiet Sun off-limb
SUMER observations by \cite{landi_etal:2002},  as these lines 
 normally show little variability.

\section{Atomic data, ion and elemental abundances}

Within the UK APAP network\footnote{www.apap-network.org}, 
we have carried out over the past few years
several large-scale structure and scattering 
$R$-matrix calculations to produce atomic data 
for dozens of ions \citep[see the review in][]{badnell_etal:2016}.
The present author has carried out a series of calculations, briefly
listed in \cite{delzanna_mason:2018}, that  have greatly 
improved the atomic data for coronal ions.
The main ions producing the majority of spectral lines in the EUV
and soft X-rays are from iron:
 \ion{Fe}{viii} \citep{delzanna_badnell:2014_fe_8},
 \ion{Fe}{ix} \citep{delzanna_etal:2014_fe_9},
 \ion{Fe}{x} \citep{delzanna_etal:12_fe_10},
\ion{Fe}{xi} \citep{delzanna_storey:2013_fe_11}, 
\ion{Fe}{xii} \citep{delzanna_etal:12_fe_12},
\ion{Fe}{xiii} \citep{delzanna_storey:12_fe_13}, and 
\ion{Fe}{xiv} \citep{delzanna_etal:2015_fe_14}.
The UK APAP  data have been included in CHIANTI Version 8 
\citep{delzanna_etal:2015_chianti_v8}.

For the present assessment, we use as a baseline CHIANTI v.8,
plus a few minor changes that are being released within  CHIANTI v.9
\citep{dere_etal:2019} 
and a large set of new atomic data that have been prepared for CHIANTI v.10
(Del Zanna et al., in preparation).
These include   APAP data for all the Be-like  
 \citep{fernandez-menchero_etal:2015_be-like} and Mg-like 
\citep{fernandez-menchero_etal:2014_mg-like} ions. New data for 
several ions have also been included:  \ion{S}{iv}  \citep{delzanna_badnell:2016_s_4},
\ion{Fe}{xiv} \citep{delzanna_etal:2015_fe_14}, \ion{Ni}{xii}
 \cite{delzanna_badnell:2016_ni_12}. 
Radiative rates from a selection of sources were also added.

\subsection{The ion charge states at zero density}

During the course of the present assessment we have tested several ion 
charge state distributions (ion abundances), calculated assuming ionization equilibrium at 
\textit{zero-density (the coronal approximation)}, i.e. assuming that 
 all the population in an ion is in the ground state.
The tables  published in CHIANTI v.6  \citep{dere_etal:09_chianti_v6}
are a significant improvement over previous ones, 
because of new ionization rates  \citep{dere:07}, 
radiative recombination rates \citep{badnell:06}, and 
dielectronic recombination rates 
(see  \citealt{badnell_etal:03} and the series of following papers).

The recombination rates for some ions were subsequently revised
in  CHIANTI v.7.1 \citep{landi_etal:12_chianti_v7.1}.
They mostly affected the charge state distribution of a few 
iron and nickel coronal ions, most notably  \ion{Fe}{viii} and \ion{Fe}{ix}.
When  using  the CHIANTI v.7.1  charge state distributions we obtained  
predicted irradiances   about a factor of two higher than 
observed, for  both  \ion{Fe}{viii} and \ion{Fe}{ix}.
It turned out that the large discrepancies were  due to  errors in the
recombination rates of  \ion{Fe}{viii} and \ion{Fe}{ix} in CHIANTI.
These were corrected and released in v.8.07, which we use here.

\subsection{Anomalous ions and  ion charge states with density effects}

The Li- and Na-like ions give rise to some of the strongest lines in the 
EUV/UV spectral region.
However, many of these ions are anomalous, in that  the emission measures obtained  
from them  are at odds with those obtained from ions of  other isoelectronic sequences.
This  was first noted by \cite{burton_etal:71}, although even earlier observations
present the same problem, as described in \cite{delzanna_mason:2018}.
 \cite{delzanna_etal:02_aumic} showed 
for the first time that the same problem is common  in stars other than 
the Sun.

The issue of anomalous ions has been largely neglected in the literature, and there is still no 
obvious explanation, although several effects  could be at play.
Departures from ionization equilibrium can enhance 
significantly some of the ions that have long ionization/recombination times,
as shown e.g.  in \cite{bradshaw_etal:04}.
Non-Maxwellian electron distributions tend to shift the 
formation temperature of the TR lines towards lower values,
leading to an enhancement, as shown for the \ion{Si}{iv} case by 
\cite{dudik_etal:2014_o_4}.

However, various  effects  related to 
the density of the plasma are always going to be present. One is 
a suppression of the dielectronic recombination, as described in 
 \cite{burgess_summers:1969}. The authors developed a 
 collisional-radiative  modelling (CRM) which was 
further improved and implemented within the Atomic Data and Analysis Structure (ADAS),
the consortium for fusion research. 
We have started a programme of new atomic calculations to model
these and other (e.g. photo-ionization) effects with up-to-date rates
 \citep{dufresne_delzanna:2018}, but for the present 
paper we have tested the  ion fractions calculated using the effective rates
as available in OPEN-ADAS\footnote{http://open.adas.ac.uk}, which include 
density effects.

The helium lines (neutral and singly-ionized) are also much stronger than 
predicted by large factors, an issue that has been discussed extensively in the 
literature.
There are  various processes that could lead to enhancements of these lines.
Recombination following photo-ionization from coronal radiation is one 
of those (see, e.g. \citealt{andretta_etal:03} and references therein).
Non-equilibrium effects such as time-dependent
ionization  (see, e.g. \citealt{bradshaw_etal:04}) and  
diffusion processes (see, e.g. \citealt{fontenla_etal:1993}),
are also important.
A recent study by \cite{golding_etal:2017} showed that time-dependent
ionization, recombination and radiative transfer effects can indeed
increase the intensities of the helium lines by a factor of 10. 
We provide no attempt here to model lines from He nor from H,
and point out that in general radiative transfer effects should be considered 
when modelling some of the lines in neutral and singly ionized atoms.

\subsection{Coronal abundances of the quiet Sun are photospheric! }

Measurements of elemental  abundances and 
temperatures  are closely linked, since the 
DEM distribution can be  used to both
describe the temperature distribution of the plasma and 
obtain relative elemental  abundances 
\citep[see the review of the methods in][]{delzanna_mason:2018}.
It is well known that solar coronal abundances present variations 
and differ from the photospheric ones, although there is 
a significant amount of controversy in the literature.  
The ratio of the coronal abundances of the 
low ($\leq$ 10 eV) first ionization potential (FIP) 
elements vs. the high-FIP ones is often  higher than the 
photospheric value (the so-called FIP bias).  
Controversy in the literature 
 regards the significant revision of the abundances of 
several important high-FIP elements (e.g. carbon, oxygen)
proposed by  \cite{asplund_etal:09}.

There is now sufficient evidence that chemical abundances in the quiet Sun transition-region
are  photospheric, while 
controversial results about the FIP bias in the corona have been published
\citep[see the reviews by][]{laming:2015,delzanna_mason:2018}.
Recent analyses of SoHO SUMER \citep{delzanna_deluca:2018}
and UVCS \citep{delzanna_etal:2018_cosie} using  CHIANTI 
v.8 of quiet Sun areas show excellent agreement with the 
\cite{asplund_etal:09} photospheric abundances, mostly 
because of the significant differences in the v.8 atomic data 
compared to earlier ones. 
We therefore use these photospheric abundances for the present analysis.
We return to the issue of chemical abundances below, when we 
present an analysis of the active Sun.

\section{The quiet Sun irradiances}

We use the PEVE spectrum as the basis for our study of the 
quiet Sun irradiances, because it was taken during the extended solar minimum.
The solar radio F10.7 flux was 69.
The Sun was mainly featureless, with a polar coronal hole  and a very small 
active region, which affected somewhat a truly `quiet Sun' measurement.

As shown in Paper V, several of the strong lines have PEVE irradiances 
that are much higher than those measured by CDS NIS, and 
also higher than the values measured in  May 2010 by the in-flight 
EVE instrument (we considered the version 5 data). 
We have therefore studied the EVE daily irradiances during 
May--June 2010 to see if a  relatively quiet period
could be found. We considered the F10.7 radio flux and 
looked at the variation 
of the EVE irradiances. We selected 2010 May 16 as one of the best dates.
Unfortunately, the Sun  
already had five small active regions on the visible side, and indeed all the 
irradiances of the lines formed above 1 MK are significantly enhanced 
in the 2010 May 16  EVE spectrum, compared to the PEVE spectrum,
despite the fact that the F10.7 radio flux was only 70.
On a side note, this comparison clearly shows that the F10.7 radio flux
is not a good indicator of very low levels of solar activity.

We have measured the irradiances of both the EVE and PEVE lines 
subtracting a background, but note that this background subtraction 
only affects the measurements of the main lines by at most 10\%,
as shown in Paper V.

The discrepancies between the 
PEVE and EVE observations are sometimes significant (50\%), well above the 
quoted uncertainties, and cannot be due to solar variability for 
lines formed below 1 MK. In fact, 
all the low-temperature emission lines (except the helium lines)
 have a very small variation during the solar cycle, as shown in 
Paper II and Paper III.
The differences  can only be due to calibration problems. 
To aid the assessment, we have considered all the historical records
of EUV irradiances, as  carried out  previously in Paper II.
We have also considered our CDS NIS irradiances during solar minimum to
assess when  the PEVE values were reasonable. In a few cases,
we have replaced the PEVE irradiances with the EVE measurements. 
The irradiances  of a selection of EUV lines  are  given in  Table~\ref{tab:lines1},
where other measurements for some of the cooler lines are also provided. 
We only list those measurements we regard as relatively accurate:
our  CDS NIS measurements of 2008 Sept 22, and the  irradiances published 
by \cite{malinovsky_heroux:73} and \cite{heroux_etal:1974}.
The  F10.7 radio flux on  2008 Sept 22 was 69.6 and
the  Sun only had a small active region, i.e. was very quiet.
\cite{malinovsky_heroux:73}  published  a calibrated  spectrum 
in the 50-300 \AA\ range with a medium resolution (0.25 \AA),
taken with  a grazing-incidence spectrometer  flown on a
rocket on 1969 April 4, when the F10.7 flux was 177.3, i.e.
when the Sun was `active', and significant contributions from 
active regions and flares were present.
\cite{heroux_etal:1974} provided irradiances also obtained from a 
sounding rocket flown in 1972 August 23 when the F10.7 flux was 120, 
i.e. the Sun was moderately active.
A more extended list of EUV 
quiet Sun irradiances is provided in  Table~\ref{tab:lines2} in the Appendix.

\subsection{DEM and predicted irradiances}

\begin{figure*}[!htbp]
\centerline{\epsfig{file=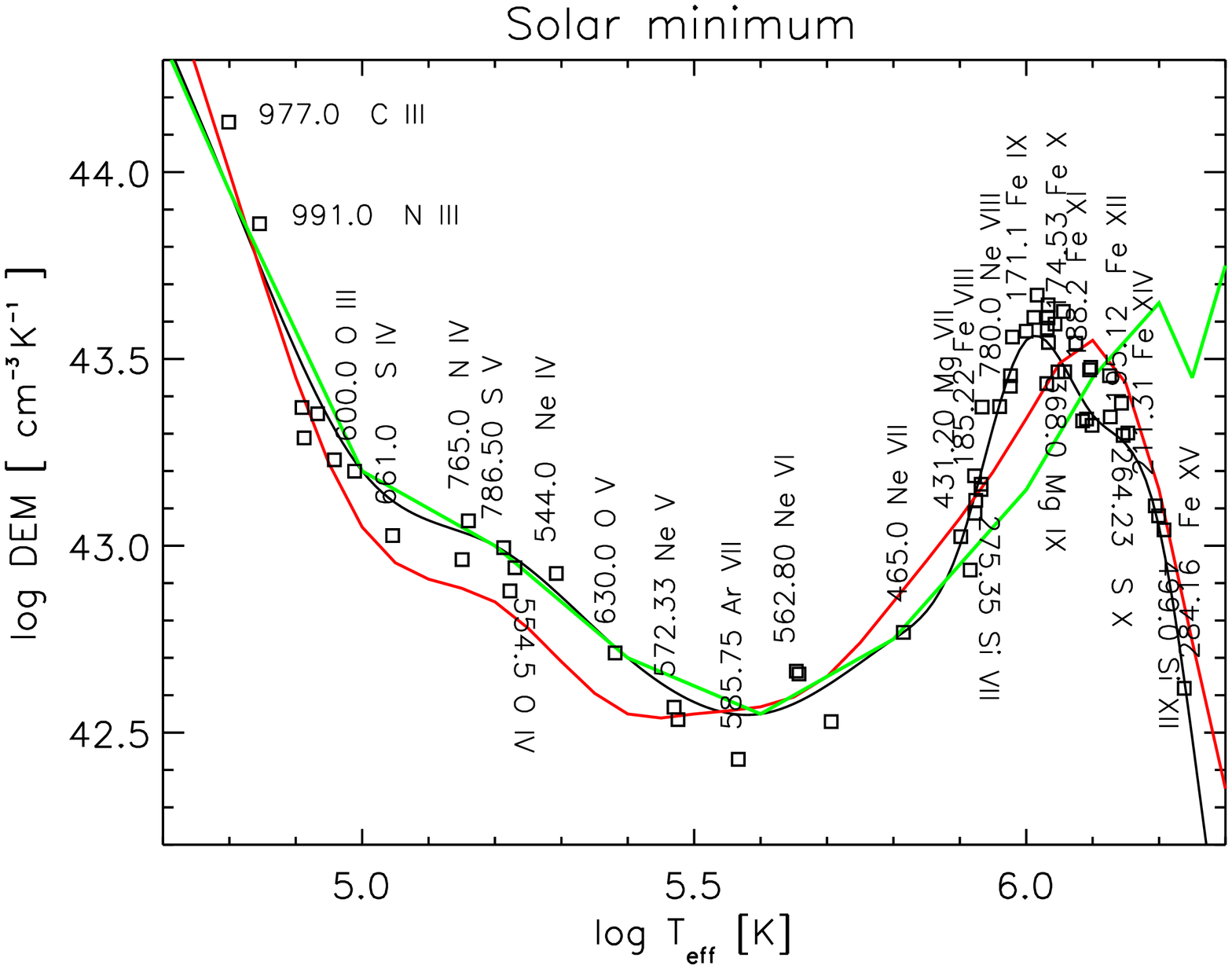, width=10.5cm,angle=90 }}
  \caption{The volume DEM distribution of the quiet Sun.
The  points indicate the ratio of the predicted
vs. observed irradiance, multiplied by the DEM value at
the effective temperature.
The labels indicate the wavelength (~\AA) and the main ion. 
The red line indicate the DEM obtained from 
the CDS NIS measurements of 2008 Sept 22, while the green line the 
DEM obtained from the active Sun using the \cite{malinovsky_heroux:73} irradiances.
}
 \label{fig:dem}
\end{figure*}

A  DEM modelling was carried out, exploring a range of parameters. 
We used the \cite{delzanna_thesis:1999} method, whereby the DEM is modeled as a spline 
distribution. 
The intensity of the  vast majority of the strongest spectral lines 
is largely independent of the density, however some transition region lines
are sensitive to the varying density in the lower part of the 
solar atmosphere. An example is the \ion{O}{v} multiplet at 760~\AA.
We adopted a model with a constant pressure 
of 5$\times$10$^{14}$ cm$^{-3}$ K$^{-1}$ and the CHIANTI v.8.07 
zero-density ion abundances. 
The resulting volume DEM is shown in Fig.~\ref{fig:dem}.
To indicate how well the DEM is constrained, we also 
plot in Fig.~\ref{fig:dem}  the ratio of the 
predicted  vs.  observed irradiance, multiplied by the DEM value
at the  effective temperature

\begin{equation}
T_{\rm eff} = {\int T~G{\left({T}\right)}~ DEM{\left({T}\right)}~dT \over 
\int G{\left({T}\right)}~DEM{\left({T}\right)}~dT}
\end{equation}
which is  an average temperature more indicative of 
where a line is formed. This is often  quite different than  
 $T_{\rm max}$, the temperature where the  $G(T)$ of a line has a maximum.

For comparison, we show in Fig.~\ref{fig:dem} the DEM obtained from the 
CDS  NIS  irradiances of 2008 Sept 22 (cf. Paper IV).
There is overall agreement, as expected.
We stress that, as noted in Paper IV, 
all the DEM distributions that we have obtained from NIS 
along the solar cycle  are all similar below 1 MK, because of the 
little variation in the irradiances of the cooler lines.

The  irradiances predicted using the DEM and the zero-density ion abundances 
are also listed in Table~\ref{tab:lines1}. 
In most of the cases, using the same DEM and the ion abundances 
calculated from OPEN-ADAS at  a constant 
pressure of 5$\times$10$^{14}$ cm$^{-3}$ K$^{-1}$, we obtain similar irradiances,
but note that the density effects tend to shift the formation temperature of a
transition-region ion  towards lower values. 
However, in a number of cases, significant discrepancies are present.
We only list the main discrepancies in Table~\ref{tab:lines1}.

\def\baselinestretch{1.}

\begin{table}[!htbp]
\caption{Observed  and predicted quiet Sun EUV  irradiances for transition-region lines,
for increased  formation temperature. } 
\begin{center}
\footnotesize
\begin{tabular}{@{}llccrlrr@{}}
\hline\hline \noalign{\smallskip}
 $\lambda_{\rm obs}$  & $I_{\rm obs}$ & $T_{\rm max}$ & $T_{\rm eff}$  & $R$ & Ion & $\lambda_{\rm exp}$ &  $r$ \\
   (\AA)             &              &  (log)       & (log)         &     &     &  (\AA)   & \\
\noalign{\smallskip}\hline\noalign{\smallskip}

1036.54 & 3.1 &  4.65 &  4.55 &  2.70 &  \ion{C}{ii} & 1036.337 & 0.33 \\ 
                                 &   &  &  &  &  \ion{C}{ii} & 1037.018 & 0.66 \\ 
         & \multicolumn{6}{l}{4.4 (EVE)}  & \\
ADAS    &     &  4.48 &  4.44 &  1.32 &  \ion{C}{ii} & &   \\

 977.04 & 51.4 &  4.94 &  4.80 &  0.66 &  \ion{C}{iii} &  977.020 & 0.99 \\ 
          & \multicolumn{6}{l}{46 (H74), 60 (EVE)}  & \\
ADAS  &  &  4.82 &  4.69 &  0.97 &  \ion{C}{iii} &  & \\

 991.60 & 3.4 &  4.94 &  4.85 &  0.77 &  \ion{N}{iii} &  991.577 & 0.87 \\ 
        & \multicolumn{6}{l}{3.9 (H74), 3.7 (EVE)}  & \\
 
 599.59 & 1.5 &  5.02 &  4.99 &  1.07 &  \ion{O}{iii} &  599.590 & 0.99 \\ 
        & \multicolumn{6}{l}{1.6 (H74), 1.2 (EVE, CDS)}  & \\

 661.40 & 0.25 &  5.06 &  5.05 &  1.24 &  \ion{S}{iv} &  661.420 & 0.88 \\ 
        & \multicolumn{6}{l}{0.37 (PEVE)}  & \\
ADAS   &    &  5.03 &  4.97 &  1.41 &  \ion{S}{iv} &   &  \\

 489.48 & 0.15 &  5.06 &  5.15 &  1.18 &  \ion{Ne}{iii} &  489.495 & 0.79 \\ 
                                 &   &  &  &  &  \ion{Ne}{iii} &  489.629 & 0.15 \\ 
ADAS     &     &  4.94 &  5.09 &  0.89 &  \ion{Ne}{iii} &   &  \\

 765.11 & 3.7 &  5.18 &  5.16 &  0.92 &  \ion{N}{iv} &  765.152 & 0.98 \\ 
        & \multicolumn{6}{l}{2.0 (H74), 2.8 (EVE)}  & \\

 786.49 & 1.5 &  5.22 &  5.21 &  0.98 &  \ion{S}{v} &  786.468 & 0.98 \\ 
         & \multicolumn{6}{l}{1.3 (H74), 1.3 (EVE)}  & \\
 ADAS    &    &   5.16 &  5.16 &  0.69 &  \ion{S}{v} &  &  \\

 554.39 & 5.5 &  5.22 &  5.23 &  1.06 &  \ion{O}{iv} &  554.076 & 0.27 \\ 
                                 &   &  &  &  &  \ion{O}{iv} &  554.514 & 0.70 \\ 
        & \multicolumn{6}{l}{6.8 (H74), 5.1 (EVE), 5.9 (CDS)}  & \\

 543.87 & 0.25 &  5.25 &  5.29 &  0.89 &  \ion{Ne}{iv} &  543.886 & 0.97 \\ 
        & \multicolumn{6}{l}{0.30 (CDS)}  & \\

 629.73 & 13.9$^*$ &  5.39 &  5.38 &  1.04 &  \ion{O}{v} &  629.732 & 0.98 \\ 
          & \multicolumn{6}{l}{12.6 (MH73), 16 (H74), 17.7 (PEVE), 8.8 (CDS)}  & \\
ADAS    &          &  5.33 &  5.32 &  0.75 &  \ion{O}{v} &   &  \\

 572.29 & 0.31 &  5.43 &  5.47 &  1.10 &  \ion{Ne}{v} &  572.336 & 0.84 \\ 
                                 &   &  &  &  &  \ion{Ne}{v} &  572.113 & 0.13 \\ 
        & \multicolumn{6}{l}{0.28 (CDS)}  & \\

 585.67 & 0.13$^*$ &  5.55 &  5.57 &  1.32 &  \ion{Ar}{vii} &  585.748 & 0.79 \\ 
         & \multicolumn{6}{l}{0.21 (PEVE)}  & \\

 562.78 & 0.61 &  5.60 &  5.65 &  0.82 &  \ion{Ne}{vi} &  562.805 & 0.81 \\ 
                                 &   &  &  &  &  \ion{Ne}{vi} &  562.711 & 0.17 \\ 
         & \multicolumn{6}{l}{0.45 (CDS)}  & \\ 

 401.85 & 0.66 &  5.61 &  5.66 &  0.84 &  \ion{Ne}{vi} &  401.941 & 0.97 \\

\noalign{\smallskip}\hline 
\end{tabular}
\end{center}
\normalsize
 \label{tab:lines1} 
\tablefoot{Each line lists the observed wavelengths 
 $\lambda_{\rm obs}$ (\AA), the measured  irradiances $I_{\rm obs}$ 
(10$^8$ photons cm$^{-2}$ s$^{-1}$ arcsec$^{-2}$),
the maximum and effective temperature 
(log values, in K;  see text)  $T_{\rm max}$ and $T_{\rm eff}$,
 the ratio $R$ between the predicted and observed values, 
the main contributing ion and CHIANTI wavelength 
$\lambda_{\rm exp}$ (\AA), and the fractional contribution  $r$
(only contributions greater than 10\% are shown) to the blend.
$I_{\rm obs}$ are prototype EVE  (PEVE) irradiances, while  
those with a $^*$ are the EVE values of 2010 May 16. Measured irradiance from 
MH73: \cite{malinovsky_heroux:73}; H74: \cite{heroux_etal:1974} and 
CDS (2008 Sept 22) are also listed in some cases.
The lines with ADAS indicate the values obtained using the OPEN-ADAS 
ion charge state distributions.
}
\end{table}

\addtocounter{table}{-1}
\begin{table}[!htbp]
\caption{Observed  and predicted quiet Sun EUV  irradiances for upper transition-region lines.}
\begin{center}
\footnotesize
\begin{tabular}{@{}llccrlrr@{}}
\hline\hline \noalign{\smallskip}
 $\lambda_{\rm obs}$  & $I_{\rm obs}$   &  $T_{\rm max}$ & $T_{\rm eff}$  & $R$ & Ion & $\lambda_{\rm exp}$   &  $r$ \\
   (\AA)             &              &  (log)       & (log)         &     &     &  (\AA)   & \\
\noalign{\smallskip}\hline\noalign{\smallskip}

 465.17 & 2.6$^*$ &  5.72 &  5.82 &  1.00 &  \ion{Ne}{vii} &  465.221 & 0.98 \\ 
         & \multicolumn{6}{l}{2.6 (H74), 3.3 (PEVE)}  & \\

ADAS    &         &  5.68 &  5.75 &  0.39 &  \ion{Ne}{vii} &  465.221 &  \\

 431.26 & 0.27 &  5.79 &  5.90 &  0.96 &  \ion{Mg}{vii} &  431.319 & 0.77 \\ 
                                 &   &  &  &  &  \ion{Mg}{vii} &  431.194 & 0.21 \\

 168.10 & 0.77 &  5.71 &  5.92 &  1.43 &  \ion{Fe}{viii} &  168.929 & 0.11 \\ 
                                 &   &  &  &  &  \ion{Fe}{viii} &  168.544 & 0.21 \\ 
                                 &   &  &  &  &  \ion{Fe}{viii} &  168.172 & 0.35 \\ 
                                 &   &  &  &  &  \ion{Fe}{viii} &  167.486 & 0.22 \\ 
         & \multicolumn{6}{l}{0.95 (MH73), 0.80 (EVE)}  & \\

 429.08 & 0.10 &  5.79 &  5.92 &  0.88 &  \ion{Mg}{vii} &  429.140 & 0.92 \\ 
 
 185.29 & 0.35 &  5.71 &  5.92 &  1.12 &  \ion{Fe}{viii} &  185.213 & 0.88 \\ 
         & \multicolumn{6}{l}{0.44 (MH73), 0.26 (EVE)}  & \\

 278.40 & 0.33 &  5.80 &  5.92 &  1.05 &  \ion{Mg}{vii} &  278.404 & 0.63 \\ 
                                 &   &  &  &  &  \ion{Si}{vii} &  278.450 & 0.30 \\ 
       & \multicolumn{6}{l}{0.34 (MH73), 0.36 (H74), 0.27 (EVE)}  & \\

 681.81 & 0.15 &  5.94 &  5.93 &  1.11 &  \ion{Na}{ix} &  681.719 & 0.78 \\ 
                                 &   &  &  &  &  \ion{S}{iii} &  681.488 & 0.18 \\ 
        & \multicolumn{6}{l}{0.25 (PEVE)}  & \\

 275.42 & 0.35 &  5.80 &  5.93 &  1.08 &  \ion{Si}{vii} &  275.361 & 0.84 \\ 
                                 &   &  &  &  &  \ion{Si}{vii} &  275.676 & 0.14 \\ 
          & \multicolumn{6}{l}{0.28 (MH73), 0.25 (H74), 0.26 (EVE)}  & \\
 ADAS   &       &  5.73 &  5.88 &  0.48 &  \ion{Si}{vii} & & \\

 466.10 & 0.45 &  5.81 &  5.93 &  0.68 &  \ion{Ca}{ix} &  466.240 & 0.98 \\ 
 
 411.15 & 0.23 &  5.87 &  5.94 &  0.53 &  \ion{Na}{viii} &  411.171 & 0.93 \\ 
 
 780.31 & 2.1 &  5.80 &  5.96 &  1.01 &  \ion{Ne}{viii} &  780.385 & 0.95 \\ 
         & \multicolumn{6}{l}{1.8 (MH73) 1.9 (H74), 2.5 (PEVE)}  & \\
 ADAS   &     &  5.74 &  5.92 &  0.79 &  \ion{Ne}{viii} &   &  \\

 315.07 & 1.1 &  5.91 &  5.98 &  1.03 &  \ion{Mg}{viii} &  315.015 & 0.97 \\ 
         & \multicolumn{6}{l}{1.35 (CDS)}  & \\ 

 430.41 & 0.75 &  5.91 &  5.98 &  0.84 &  \ion{Mg}{viii} &  430.454 & 0.97 \\

 170.98 & 7.0 &  5.91 &  6.00 &  0.95 &  \ion{Fe}{ix} &  171.073 & 0.99 \\ 
         & \multicolumn{6}{l}{4.4 (MH73), 4.2 (H74), 3.95 (EVE)}  & \\
 
 319.86 & 1.4 &  5.94 &  6.01 &  0.89 &  \ion{Si}{viii} &  319.840 & 0.98 \\ 
        & \multicolumn{6}{l}{1.38  (CDS)}  & \\ 
 ADAS   &     &  5.87 &  5.97 &  0.66 &  \ion{Si}{viii} &   &  \\

 706.05 & 0.52 &  5.99 &  6.02 &  0.78 &  \ion{Mg}{ix} &  706.060 & 0.95 \\ 
       & \multicolumn{6}{l}{0.34 (EVE)}  & \\
 
 557.60 & 0.26 &  5.89 &  6.03 &  1.30 &  \ion{Ca}{x} &  557.766 & 0.99 \\ 
        & \multicolumn{6}{l}{0.21 (CDS)}  & \\ 
 
 368.10 & 5.8$^*$ &  5.99 &  6.03 &  0.86 &  \ion{Mg}{ix} &  368.071 & 0.96 \\ 
       & \multicolumn{6}{l}{6.2 (H74) 8.1 (PEVE), 5.8 (CDS)}  & \\

 272.05 & 0.42 &  5.28 &  6.03 &  0.79 &  \ion{Si}{x} &  271.992 & 0.75 \\ 

 443.80 & 0.25 &  5.99 &  6.03 &  1.00 &  \ion{Mg}{ix} &  443.404 & 0.16 \\ 
                                 &   &  &  &  &  \ion{Mg}{ix} &  443.973 & 0.78 \\ 
 
 198.51 & 0.26 &  5.96 &  6.04 &  0.84 &  \ion{S}{viii} &  198.554 & 0.65 \\ 
                                 &   &  &  &  &  \ion{Fe}{xi} &  198.538 & 0.27 \\ 
          & \multicolumn{6}{l}{0.23 (MH73), 0.24 (EVE)}  & \\

\noalign{\smallskip}\hline 
\end{tabular}
\end{center}
\normalsize
\end{table}


\addtocounter{table}{-1}
\begin{table}[!htbp]
\caption{Observed  and predicted quiet Sun EUV  irradiances for coronal lines.}
\begin{center}
\footnotesize
\begin{tabular}{@{}llccrlrr@{}}
\hline\hline \noalign{\smallskip}
 $\lambda_{\rm obs}$  & $I_{\rm obs}$   &  $T_{\rm max}$ & $T_{\rm eff}$  & $R$ & Ion & $\lambda_{\rm exp}$   &  $r$ \\
   (\AA)             &              &  (log)       & (log)         &     &     &  (\AA)   & \\
\noalign{\smallskip}\hline\noalign{\smallskip}

 345.13 & 0.90 &  6.06 &  6.06 &  0.72 &  \ion{Si}{ix} &  345.121 & 0.74 \\ 
                                 &   &  &  &  &  \ion{Si}{ix} &  344.954 & 0.21 \\ 
        & \multicolumn{6}{l}{1.37 (CDS)}  & \\

 174.57 & 3.7 &  6.05 &  6.06 &  1.03 &  \ion{Fe}{x} &  174.531 & 0.98 \\ 
         & \multicolumn{6}{l}{4.6 (MH73), 4.1 (H74), 3.8 (EVE)}  & \\
 
 624.94 & 2.1 &  6.07 &  6.07 &  0.76 &  \ion{Mg}{x} &  624.968 & 0.92 \\ 
         & \multicolumn{6}{l}{1.4 (CDS)}  & \\

 148.38 & 0.38 &  6.11 &  6.08 &  1.14 &  \ion{Ni}{xi} &  148.377 & 0.99 \\ 
        & \multicolumn{6}{l}{0.59 (H74), 0.45 (EVE)}  & \\
 
  188.29 & 2.3 &  6.13 &  6.09 &  1.08 &  \ion{Fe}{xi} &  188.216 & 0.50 \\ 
                                 &   &  &  &  &  \ion{Fe}{xi} &  188.299 & 0.31 \\ 
                                 &   &  &  &  &  \ion{Fe}{ix} &  188.493 & 0.13 \\ 
 
 178.09 & 0.14 &  6.12 &  6.10 &  0.78 &  \ion{Fe}{xi} &  178.058 & 0.91 \\

 180.41 & 2.7 &  6.13 &  6.10 &  1.07 &  \ion{Fe}{xi} &  180.401 & 0.84 \\ 
 
 264.30 & 0.38 &  6.18 &  6.13 &  0.71 &  \ion{S}{x} &  264.231 & 0.98 \\

 152.05 & 0.11 &  6.24 &  6.14 &  0.80 &  \ion{Ni}{xii} &  152.151 & 0.94 \\ 
        & \multicolumn{6}{l}{0.36  (H74), 0.23 (EVE)}  & \\
 
 195.09 & 1.4 &  6.20 &  6.15 &  0.96 &  \ion{Fe}{xii} &  195.119 & 0.95 \\ 
 
 202.10 & 0.96 &  6.25 &  6.15 &  0.92 &  \ion{Fe}{xiii} &  202.044 & 0.64 \\ 
                                 &   &  &  &  &  \ion{Fe}{xi} &  202.424 & 0.12 \\ 
 
 211.40 & 0.33 &  6.29 &  6.19 &  0.97 &  \ion{Fe}{xiv} &  211.317 & 0.75 \\ 
                                 &   &  &  &  &  \ion{Ni}{xi} &  211.430 & 0.12 \\ 
 
 499.41 & 0.34 &  6.29 &  6.20 &  0.95 &  \ion{Si}{xii} &  499.406 & 0.94 \\ 
ADAS    &      &  6.24 &  6.18 &  2.14 &  \ion{Si}{xii} &   &  \\

 284.12 & 0.46 &  6.34 &  6.24 &  1.05 &  \ion{Al}{ix} &  284.042 & 0.20 \\ 
                                 &   &  &  &  &  \ion{Fe}{xv} &  284.163 & 0.76 \\

\noalign{\smallskip}\hline 
\end{tabular}
\end{center}
\normalsize
\end{table}


The present results in terms of anomalous ions 
differ in some respect with  the earlier ones. 
In fact, the irradiances of  the higher-temperature lines,
the Li-like \ion{Ne}{viii}, \ion{Na}{ix}, \ion{Mg}{x}, 
\ion{Si}{xii} are well represented, unlike many previous reports. 
The reasons  could be due to the better instrument 
calibration, and the fact that we are analysing the solar irradiance
during the solar minimum. When active regions are present, 
non-equilibrium effects are likely to become stronger.

For some low-temperature ions such as \ion{C}{ii} and \ion{C}{iii}
(as well as the anomalous ion \ion{C}{iv}, not shown here),   the OPEN-ADAS rates 
provide a significant improvement over the zero-density ones.
However,  in many  cases significant discrepancies are present.
The most obvious ones are those for the strong \ion{Ne}{vii}, 
 \ion{Si}{vii}, and \ion{Si}{xii} where the predicted irradiances are
off by a factor of two. 
The problems are present also if the ADAS data are used to infer 
a DEM. 
It is impossible  to ascertain the reasons for such discrepancies, given that 
information on the basic rates used in the ADAS CRM is not available.

\subsection{Quiet Sun EUV spectra}

\begin{figure*}[!htbp]
\centerline{\epsfig{file=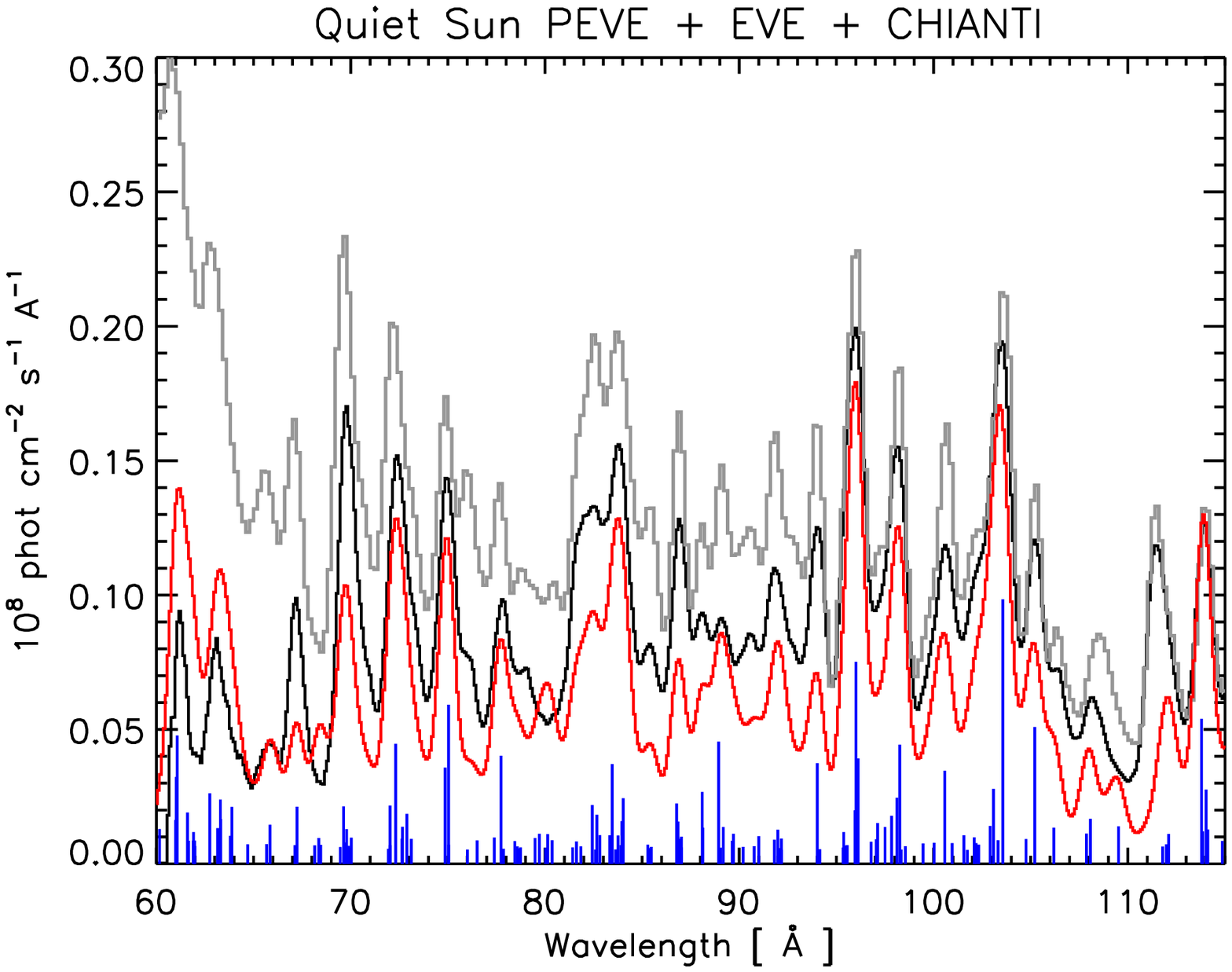, width=7.5cm,angle=0 }
\epsfig{file=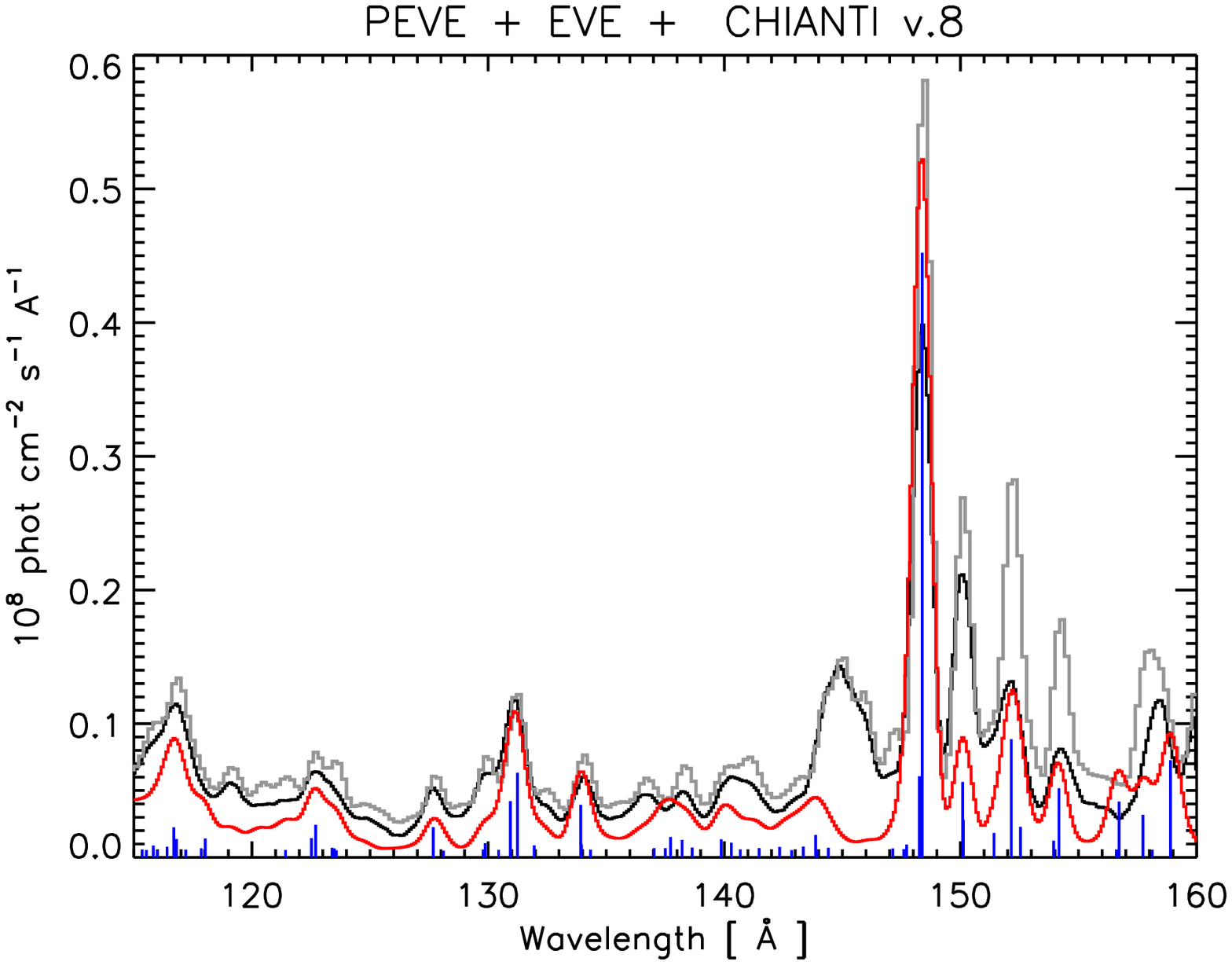, width=7.5cm,angle=0 }}
\centerline{\epsfig{file=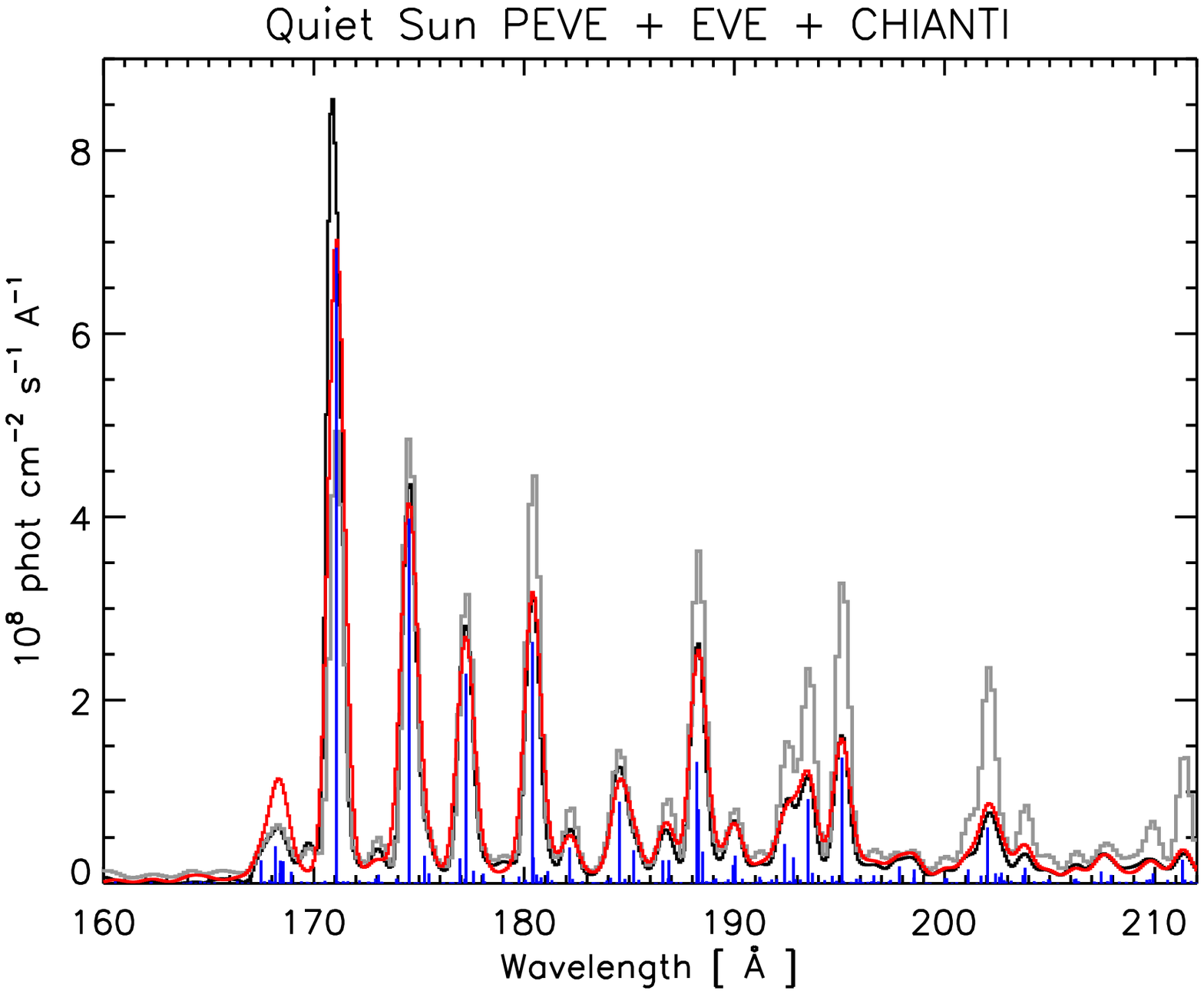, width=7.5cm,angle=0 }
\epsfig{file=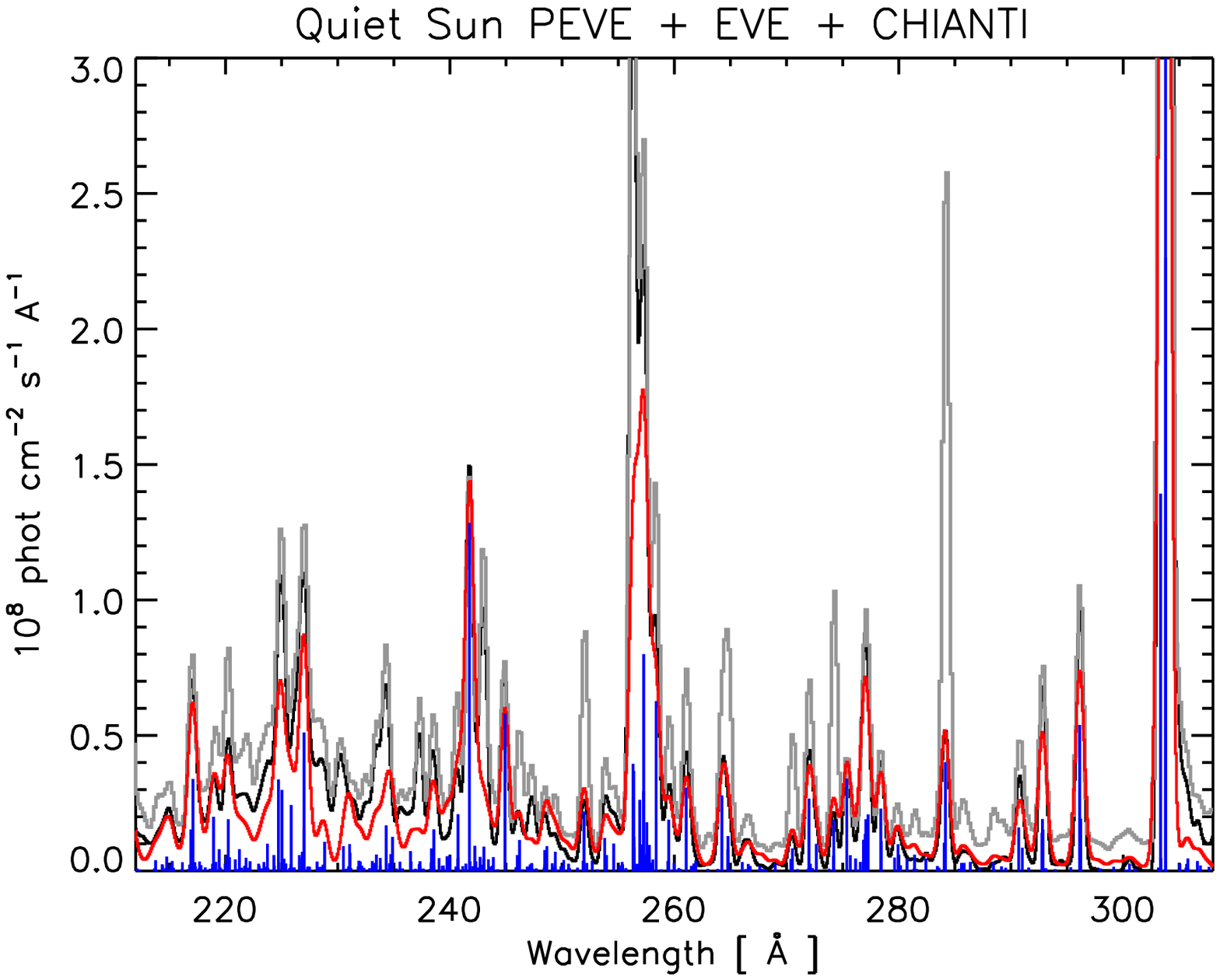, width=7.5cm,angle=0 }}
\centerline{\epsfig{file=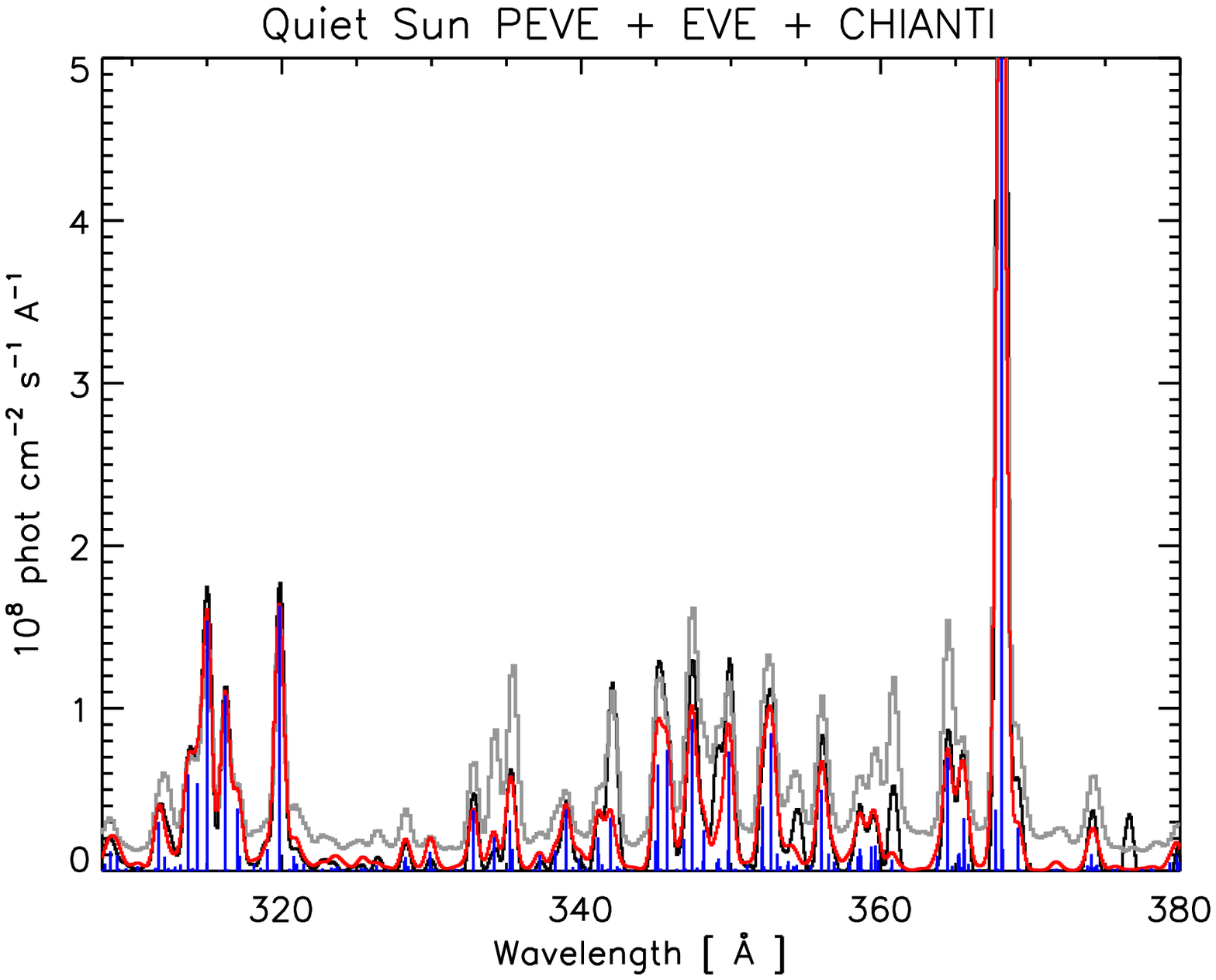, width=7.5cm,angle=0 }
\epsfig{file=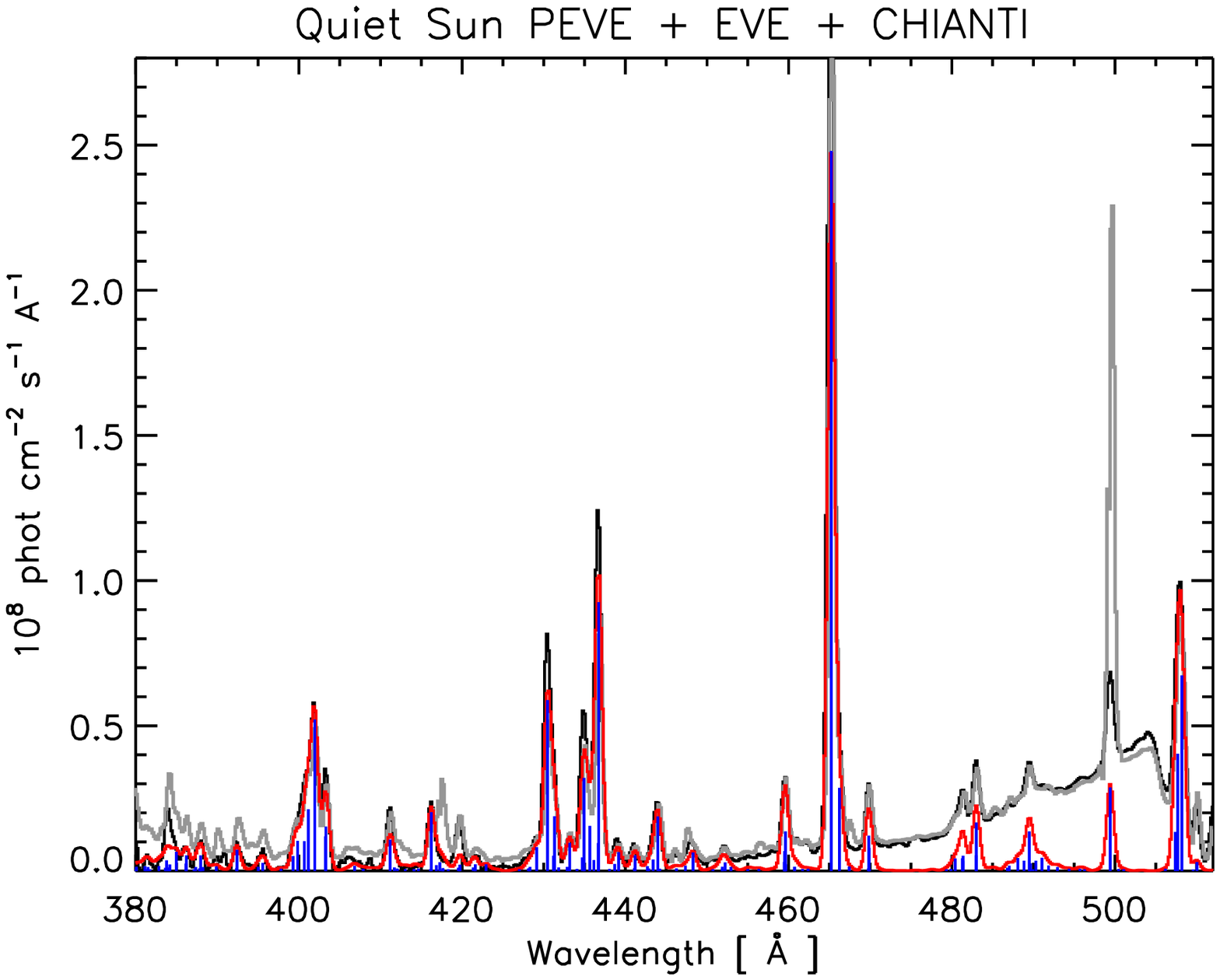, width=7.5cm,angle=0 }}
\centerline{\epsfig{file=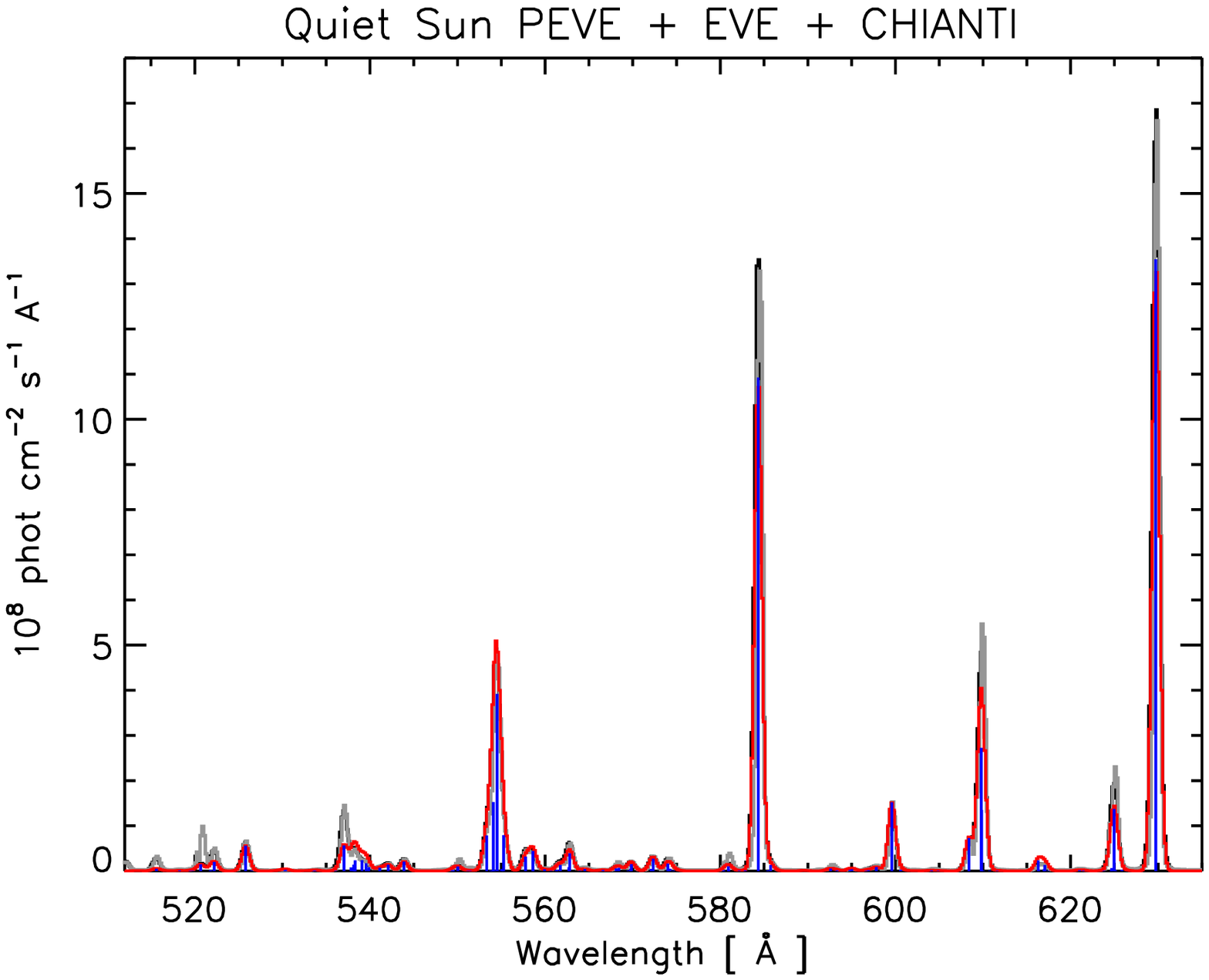, width=7.5cm,angle=0 }
\epsfig{file=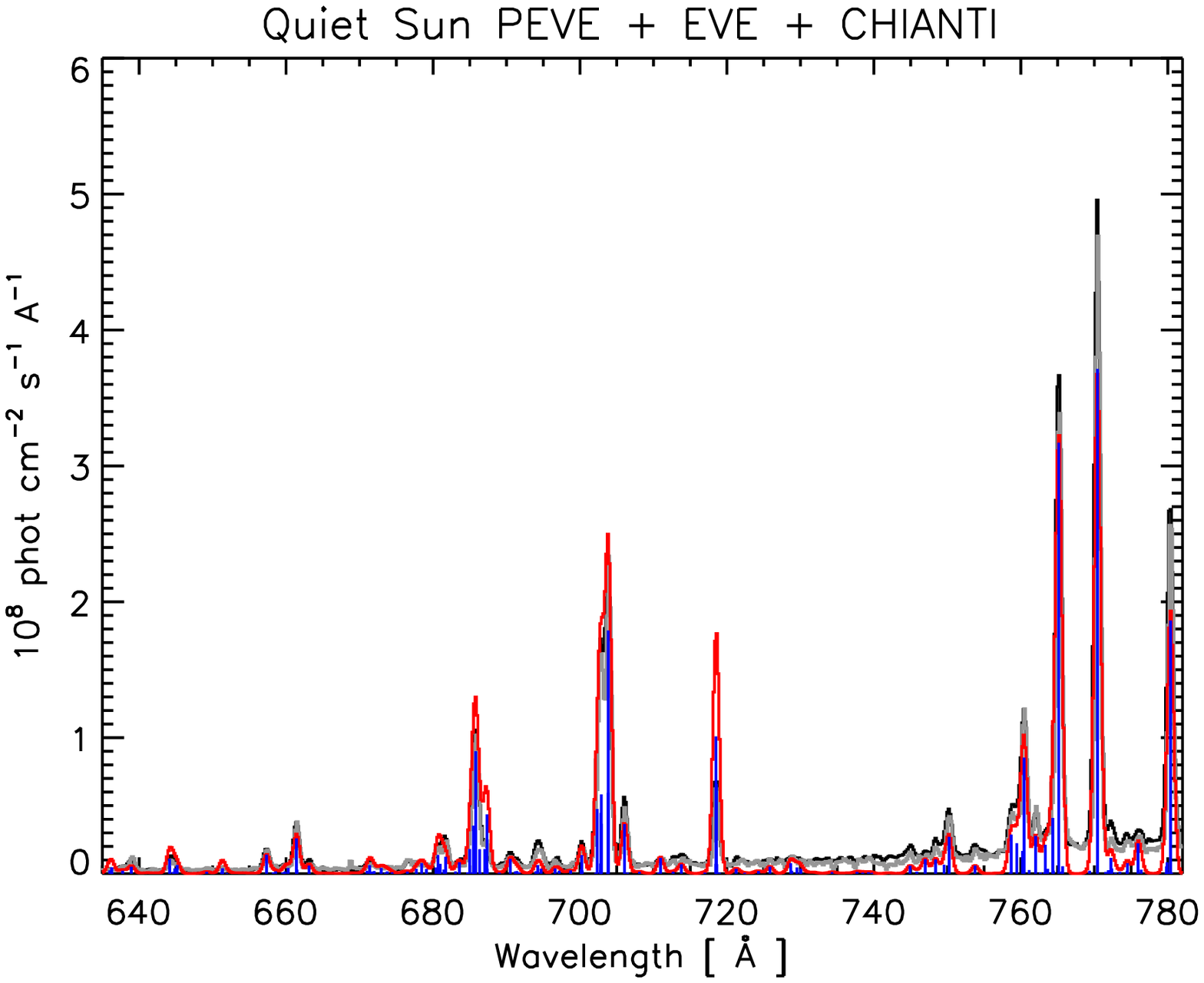, width=7.5cm,angle=0 }}
  \caption{Observed PEVE quiet Sun spectrum in 2008 (black), with the 
observed EVE spectrum on 2010 May 16 (grey) and the 
predicted spectrum (red). The positions  and intensities of the main contributing
lines are  shown by vertical blue lines.  }
  \label{fig:simul1}
\end{figure*}

The DEM and the same set of parameters used for the inversion 
were then used to produce a modeled spectrum, to be compared to the observed one,
to assess for completeness in the atomic data.
The PEVE quiet Sun spectrum of  2008 is shown in Fig.~\ref{fig:simul1} in black,
while the quiet Sun EVE spectrum on 2010 May 16 is in grey.
Fig.~\ref{fig:simul1} shows that any small solar activity affects
significantly the solar spectrum at most  wavelengths.
The predicted line profiles are approximated with Gaussians and shown in red
in Fig.~\ref{fig:simul1}, while 
the positions  and intensities of the main contributing
lines are  shown by vertical blue lines. 
We recall that several PEVE irradiances are not correct, and  that 
lines from H and He are notoriously difficult to model. 
We now briefly summarise the main findings for each wavelength region.

\subsubsection{The soft X-rays: 60--160~\AA}

As shown in Fig.~\ref{fig:simul1}, relatively good agreement 
with the PEVE data is found, although further work is still needed 
in this relatively unexplored spectral region.
When calculating the atomic data for the coronal iron ions, 
we were surprised to find  that many lines have increased intensities, when compared
to distorted wave calculations (available since CHIANTI v.7.1, see 
\citealt{landi_etal:12_chianti_v7.1}). This was due to 
 resonance enhancements in the excitation rates and cascading effects.
These new atomic data and identifications have been included in 
CHIANTI v.8 and allowed the first identifications of  the strongest  
iron coronal lines in this spectral region, using 
much higher-resolution solar and laboratory  spectra  \citep{delzanna:12_sxr1}.

\subsubsection{The 160--310~\AA\ spectral region}

Most of the 160--290~\AA\ spectral region is now  well understood,
after several studies based on the Hinode EIS spectra
\citep[cf.][]{delzanna:12_atlas}, and the new atomic data previously 
mentioned.
The agreement between theory and observations is within a few percent, 
with the exception of one of the \ion{Fe}{viii} lines at 168.7\AA\ and 
the strong \ion{Fe}{ix} resonance line at 171~\AA.
As we have previously mentioned, the disagreement is much larger 
when considering CHIANTI ion abundances earlier than 8.0.7. 

We note that the  PEVE irradiances of the \ion{Fe}{viii} lines around 168\AA\ are in 
good agreement with the EVE and  \cite{malinovsky_heroux:73}
ones and  several 
weaker  \ion{Fe}{viii} lines at longer wavelengths are very well reproduced. 
All the \ion{Fe}{viii} EUV lines were benchmarked against high-resolution 
solar and laboratory observations \citep{delzanna:09_fe_8}, showing good agreement.

We also note an inconsistency in the measured irradiances of the 
 \ion{Fe}{ix} 171~\AA\ line. 
As shown in  Table~\ref{tab:lines1}, the PEVE irradiance 
 is 7.0 (10$^8$ photons cm$^{-2}$ s$^{-1}$ arcsec$^{-2}$)
while the EVE v.5 value is much lower, 3.95. 
The \cite{malinovsky_heroux:73} value is 4.4. Excellent agreement with theory 
is found  when using the  PEVE value. We note that all the other 
 \ion{Fe}{ix} lines at longer wavelengths,  most notably the 
two density-sensitive lines at 241.74 and 244.91~\AA, are also very well 
reproduced.

In an earlier benchmark \citep{delzanna_etal:2011_aia}, we showed 
that several coronal lines in the 205--240~\AA\ are still 
unidentified, so further work is needed in this spectral region.
This can also be seen looking at  Table~\ref{tab:lines2}, although  
 these missing lines are not very strong in solar irradiance spectra.
Aside from this region, good agreement is found, as  Fig.~\ref{fig:simul1} shows.
One exception is the series of \ion{He}{ii} lines at 304, 256, 243~\AA.

\subsubsection{The 310-380~\AA\ spectral region}

The 310-380~\AA\ spectral region is also well reproduced, 
as significant effort in improving the atomic data was produced before
the launch of SoHO, as shown by 
the earlier  atomic  benchmark studies using the 
SoHO CDS NIS 1 spectra \citep{delzanna_thesis:1999}.
Many of the lines in this spectral region  are due to coronal iron ions, 
for which the new atomic data in CHIANTI v.8 are a significant improvement.
For several iron lines in this band enhancements of about 30\% are present. 
This is caused by resonance enhancement and cascading effects
increasing the decays from the lower levels producing the lines in 
this spectral region.

\subsubsection{The 380--510~\AA\ spectral region}

The quiet Sun spectrum in the 380--510~\AA\  region 
is dominated by transition-region lines from Ne and Mg ions.
This spectral region is also well understood, because 
various studies based on the SoHO CDS GIS spectra were 
performed \citep{delzanna_thesis:1999}, and because the atomic data
for the ions in this spectral range are relatively easier to calculate.
 Fig.~\ref{fig:simul1} shows that there is very good agreement
between  observed and predicted irradiances. 
We note that the neutral helium continuum (with the edge at 504~\AA) 
cannot be  modeled properly without taking into account 
radiative transfer effects.

\subsubsection{The 510-630~\AA\ spectral region}

The 510-630~\AA\ spectral region is also well known and 
complete in terms of atomic data,
 since SoHO CDS NIS 2 spectra have been used to benchmark 
the atomic data \citep{delzanna_thesis:1999}, and most of the ions
are low charge. 
Very good agreement is found, with the exception of the 
series of \ion{He}{i} lines.

\subsubsection{The 630-780~\AA\ spectral region}

The 630-780~\AA\ spectral region has been studied 
 using CDS GIS spectra \citep{delzanna_thesis:1999}.
Good overall agreement between observed and predicted line irradiances is 
found. 
We note that the intensities of the \ion{O}{ii} 718~\AA\ line and of the  
\ion{O}{v} multiplet at 760~\AA\ are very sensitive to the 
temperatures and densities of the model.
With the present  atomic data, the \ion{O}{v} multiplet is well represented,
but the \ion{O}{ii} line is not. 
Using the ADAS  ion abundances for \ion{O}{ii} does not improve
the comparison.
We also note that the hydrogen continuum (with the edge at 912~\AA) 
cannot be  modeled properly without taking into account 
radiative transfer effects.

\subsubsection{The 780--912~\AA\ spectral region}

The few  \ion{O}{iii} and \ion{O}{iv} in this wavelength region 
are well represented (cf.  Table~\ref{tab:lines2}), 
but all the  \ion{O}{ii} lines are incorrectly modeled
(also using the ADAS  ion abundances).

\subsubsection{The 912--1045~\AA\ spectral region}

The \ion{C}{iii} resonance line at 977~\AA\ is relatively well represented
with the CHIANTI v.8.0.7 ion fractions, although the ADAS ion abundances
improve agreement.
The  \ion{N}{iii} 991.57~\AA\ line is also well represented 
(cf.  Table~\ref{tab:lines2}). 
The doublet at 1031.9, 1037.6~\AA\ from the anomalous ion 
\ion{O}{vi}  is under-predicted by more than a factor of 2.
Only a small improvement is obtained using the ADAS ion abundances. 



\section{The active Sun irradiances and chemical abundances}

\begin{figure}[!htbp]
\centerline{\epsfig{file=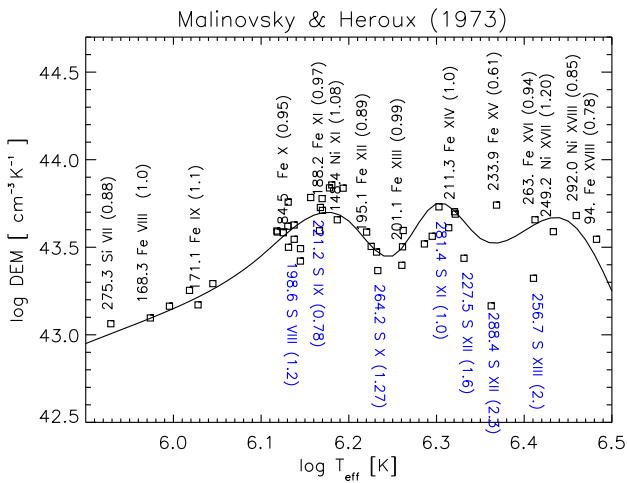, width=6.5cm,angle=90 }}
  \caption{The DEM distribution obtained from the 
active Sun  \cite{malinovsky_heroux:73} irradiances.
The  points indicate the ratio of the predicted
vs. observed irradiance, multiplied by the DEM value at
the effective temperature. The labels indicate the wavelength (\AA),
main ion, and  the ratio of the predicted vs. observed irradiance (in brackets). }
 \label{fig:malinovski_dem}
\end{figure}

The \cite{laming_etal:1995} study, based on the  \cite{malinovsky_heroux:73} spectrum,
is often cited in the literature to support the argument that 
the solar corona has a FIP bias of about 4.
However, \cite{laming_etal:1995} clearly stated that 
 nearly  photospheric abundances were found  when considering 
lines formed in the transition region, up to 1 MK.
The FIP bias of 3--4 was obtained from  higher-temperature lines.
Clearly, this hot emission would be significantly biased by the 
active regions  and not the quiet Sun, where 
almost no emission above 1 MK is present.

So,  to interpret the full-Sun spectra such as those of 
\cite{malinovsky_heroux:73} we need to have a good understanding of the 
abundances in active regions. 
Controversial results about the FIP bias in the 
hot (3 MK) cores of active regions have  been published in the literature.
However, a recent revision of EUV (Hinode/EIS)
and X-ray (SMM/FCS)  measurements of several active regions 
have indicated a remarkable consistency, with an FIP bias of 
about 3.2 \citep{delzanna:2013_multithermal,delzanna_mason:2014}.
\cite{delzanna:2013_multithermal} also showed that 
it must be the low-FIP elements that are
 enhanced by at least a  factor of 3, compared to the photospheric 
values. 
Note that the \cite{delzanna:2013_multithermal} results were obtained 
using  a new  Hinode EIS calibration \citep{delzanna:13_eis_calib}.

Considering the significant updates in atomic data since  
the \cite{laming_etal:1995} study, and the 
changes in the photospheric abundances recommended by \cite{asplund_etal:09}, 
 it is worth reconsidering the well-calibrated  \cite{malinovsky_heroux:73} EUV
irradiances.
We have chosen a selection of lines in the 
 \cite{malinovsky_heroux:73} EUV spectrum, avoiding the lines strongly 
density-dependent. We note that in several cases, the authors
had deblended the intensities of lines close in wavelength.
The spectrum must have been obtained after or during flaring emission,
as the \ion{Fe}{xviii} line at 93.9~\AA\ was relatively strong.
We have taken the deblended value as recommended by \cite{malinovsky_heroux:73}
but applied a correction factor of 2 to its irradiance, following 
the problems in the soft X-rays  described in \cite{delzanna:12_sxr1}.
With this correction, good agreement between predicted and observed
irradiance is obtained. 

We used the set of atomic data previously
described, the CHIANTI v.8.07 zero-density
ion abundances,  and the \cite{asplund_etal:09} photospheric abundances. 
We assumed  a constant density of 1$\times$10$^{9}$ cm$^{-3}$ for the 
calculation of the line emissivities.

The DEM,  shown in  Fig.~\ref{fig:malinovski_dem}, is significantly different 
than the quiet Sun one for temperatures higher than 1 MK, as expected.
The  predicted irradiances  are shown in Table~\ref{tab:mh73}.
The analysis confirms our above-mentioned  results, in particular the 
fact that up to 1.5 MK the chemical abundances are nearly photospheric
(within 30\%, cf.  \ion{S}{viii}, \ion{S}{ix}, \ion{S}{x},
and \ion{S}{xi} vs. the low-FIP iron ions).
The \ion{S}{x} abundance is well constrained as the 264.2~\AA\ line 
is relatively strong and strong iron lines are formed at the same temperature. 

We clearly see, however, the effect of the presence of 
active regions on this active Sun irradiances. 
The iron/nickel  lines require two secondary  peaks at higher temperatures. 
The abundance of Sulphur deviates by a factor of 
about 2  around 2 MK, as seen in  \ion{S}{xii} and  \ion{S}{xiii}.
The  \ion{S}{xii} abundance is well constrained relative to the
 \ion{Fe}{xiv} resonance line at 211.3~\AA, which is formed at similar temperatures. 
The  \ion{S}{xiii} abundance is well constrained relative to the
 \ion{Fe}{xvi} emission. 
 
These results are  in excellent 
agreement with our active region Hinode EIS results \citep{delzanna:12_atlas},
but not with those of \cite{brooks_etal:2015} nor with the previous 
analysis of \cite{laming_etal:1995}. 

Finally, we note that the DEM  shown in  Fig.~\ref{fig:malinovski_dem} is actually 
misleading, as we obtained it assuming constant abundances of the 
low-FIP  iron/nickel elements. Consequently, the Sulphur abundance at 2 MK 
appears 2 times lower than photospheric.

\begin{table}[!htbp]
\caption{Observed  and predicted EUV  irradiances from the 
 \cite{malinovsky_heroux:73} spectrum (same as  Table~\ref{tab:lines1}). } 
\begin{center}
\footnotesize
\begin{tabular}{@{}llccrlrr@{}}
\hline\hline \noalign{\smallskip}
 $\lambda_{\rm obs}$  & $I_{\rm obs}$   &  $T_{\rm max}$ & $T_{\rm eff}$  & $R$ & Ion & $\lambda_{\rm exp}$  & $r$ \\
\noalign{\smallskip}\hline\noalign{\smallskip}

 629.73 & 12.7 &  5.39 &  5.38 &  1.14 &  \ion{O}{v} &  629.732 & 0.98 \\ 
 
 275.42 & 0.28 &  5.79 &  5.93 &  0.88 &  \ion{Si}{vii} &  275.361 & 0.98 \\ 
 
 168.10 & 0.96 &  5.71 &  5.97 &  1.00 &  \ion{Fe}{viii} &  168.544 & 0.19 \\ 
                                 &   &  &  &  &  \ion{Fe}{viii} &  168.172 & 0.31 \\ 
                                 &   &  &  &  &  \ion{Fe}{viii} &  167.486 & 0.19 \\

 186.60 & 0.25 &  5.71 &  6.02 &  0.87 &  \ion{Ca}{xiv} &  186.610 & 0.11 \\ 
                                 &   &  &  &  &  \ion{Fe}{viii} &  186.598 & 0.86 \\ 
 
 170.98 & 4.4 &  5.90 &  6.03 &  1.11 &  \ion{Fe}{ix} &  171.073 & 0.99 \\ 
 
 278.40 & 0.34 &  5.79 &  6.04 &  0.93 &  \ion{P}{xii} &  278.286 & 0.17 \\ 
                                 &   &  &  &  &  \ion{Mg}{vii} &  278.404 & 0.52 \\ 
                                 &   &  &  &  &  \ion{Si}{vii} &  278.450 & 0.25 \\ 
 
 174.57 & 4.6 &  6.05 &  6.12 &  0.87 &  \ion{Fe}{x} &  174.531 & 0.98 \\ 
 
 177.20 & 2.6 &  6.05 &  6.12 &  0.88 &  \ion{Fe}{x} &  177.240 & 0.98 \\ 
 
 184.46 & 1.0 &  6.05 &  6.13 &  0.95 &  \ion{Fe}{x} &  184.537 & 0.93 \\ 
 
 296.15 & 0.94 &  6.05 &  6.13 &  0.92 &  \ion{Si}{ix} &  296.211 & 0.23 \\ 
                                 &   &  &  &  &  \ion{Si}{ix} &  296.113 & 0.74 \\ 
 
 225.00 & 0.51 &  6.06 &  6.13 &  0.67 &  \ion{Si}{ix} &  225.025 & 0.98 \\ 
 
 198.51 & 0.23 &  5.96 &  6.13 &  1.22 &  \ion{S}{viii} &  198.554 & 0.43 \\ 
                                 &   &  &  &  &  \ion{Fe}{xi} &  198.538 & 0.48 \\ 
 
 185.29 & 0.44 &  5.72 &  6.14 &  0.97 &  \ion{Ni}{xvi} &  185.230 & 0.29 \\ 
                                 &   &  &  &  &  \ion{Fe}{viii} &  185.213 & 0.63 \\ 
 
 609.90 & 6.0 &  5.21 &  6.14 &  1.17 &  \ion{Mg}{x} &  609.794 & 0.79 \\ 
                                 &   &  &  &  &  \ion{O}{iv} &  609.830 & 0.20 \\ 
 
 770.41 & 3.0 &  5.80 &  6.14 &  1.66 &  \ion{Ne}{viii} &  770.428 & 1.00 \\ 
 
 780.31 & 1.8 &  5.80 &  6.14 &  1.41 &  \ion{Ne}{viii} &  780.385 & 1.00 \\ 
 
 221.24 & 0.14 &  6.06 &  6.16 &  0.78 &  \ion{S}{ix} &  221.241 & 0.98 \\ 
 
 178.09 & 0.18 &  6.13 &  6.17 &  1.26 &  \ion{Fe}{xi} &  178.058 & 0.95 \\ 
 
 180.41 & 5.3 &  6.13 &  6.17 &  0.93 &  \ion{Fe}{xi} &  180.401 & 0.87 \\ 
 
 272.05 & 0.75 &  5.29 &  6.17 &  0.83 &  \ion{Si}{x} &  271.992 & 0.84 \\ 
 
 188.29 & 3.8 &  6.13 &  6.17 &  0.97 &  \ion{Fe}{xi} &  188.216 & 0.60 \\ 
                                 &   &  &  &  &  \ion{Fe}{xi} &  188.299 & 0.36 \\ 
 
 290.70 & 0.31 &  6.05 &  6.18 &  0.73 &  \ion{Si}{ix} &  290.687 & 0.75 \\ 
                                 &   &  &  &  &  \ion{Fe}{xiv} &  290.749 & 0.15 \\ 
 
 223.70 & 0.27 &  6.07 &  6.18 &  0.70 &  \ion{Si}{ix} &  223.744 & 0.60 \\ 
                                 &   &  &  &  &  \ion{Fe}{xiii} &  223.778 & 0.21 \\ 
 
 148.38 & 0.72 &  6.11 &  6.19 &  1.08 &  \ion{Ni}{xi} &  148.377 & 0.99 \\ 
 
 261.06 & 0.89 &  6.15 &  6.19 &  0.69 &  \ion{Si}{x} &  261.056 & 0.97 \\ 
 
 195.09 & 4.8 &  6.19 &  6.22 &  0.89 &  \ion{Fe}{xii} &  195.119 & 0.92 \\ 
 
 624.94 & 2.8 &  6.06 &  6.23 &  1.00 &  \ion{Mg}{x} &  624.968 & 0.99 \\ 
 
 259.50 & 0.68 &  6.18 &  6.23 &  1.01 &  \ion{S}{x} &  259.497 & 0.80 \\ 
                                 &   &  &  &  &  \ion{Fe}{xii} &  259.418 & 0.15 \\ 
 
 264.20 & 0.65 &  6.18 &  6.23 &  1.27 &  \ion{S}{x} &  264.231 & 0.98 \\ 
 
 239.80 & 0.17 &  6.19 &  6.26 &  1.25 &  \ion{S}{xi} &  239.817 & 0.63 \\ 
                                 &   &  &  &  &  \ion{Fe}{xi} &  239.780 & 0.31 \\ 
 
 201.10 & 1.1 &  6.24 &  6.26 &  0.99 &  \ion{Fe}{xiii} &  201.126 & 0.84 \\ 
                                 &   &  &  &  &  \ion{Fe}{xi} &  201.112 & 0.12 \\ 
 
 152.05 & 0.51 &  6.24 &  6.26 &  0.81 &  \ion{Ni}{xii} &  152.151 & 0.95 \\ 
 
 215.16 & 0.21 &  5.42 &  6.29 &  1.47 &  \ion{O}{v} &  215.245 & 0.11 \\ 
                                 &   &  &  &  &  \ion{S}{xii} &  215.167 & 0.80 \\ 
 
 285.80 & 0.23 &  6.27 &  6.30 &  1.49 &  \ion{S}{xi} &  285.823 & 0.96 \\ 
 
 281.40 & 0.32 &  6.28 &  6.30 &  1.05 &  \ion{S}{xi} &  281.402 & 0.97 \\ 
 
 157.75 & 0.30 &  6.30 &  6.31 &  1.33 &  \ion{Ni}{xiii} &  157.729 & 0.91 \\

 164.13 & 0.24 &  6.31 &  6.32 &  1.01 &  \ion{Ni}{xiii} &  164.150 & 0.91 \\ 
 
 211.40 & 3.9 &  6.29 &  6.32 &  1.03 &  \ion{Fe}{xiv} &  211.317 & 0.95 \\ 
 
 227.50 & 0.30 &  5.43 &  6.33 &  1.62 &  \ion{S}{xii} &  227.490 & 0.83 \\ 
 
 288.35 & 0.35 &  6.35 &  6.36 &  2.29 &  \ion{S}{xii} &  288.434 & 0.98 \\ 
 
 233.90 & 0.42 &  6.34 &  6.37 &  0.61 &  \ion{Fe}{xv} &  233.866 & 0.88 \\

 256.70 & 1.1 &  6.42 &  6.41 &  2.01 &  \ion{S}{xiii} &  256.685 & 0.97 \\ 
 
 263.05 & 0.73 &  6.43 &  6.41 &  0.94 &  \ion{Fe}{xvi} &  262.976 & 0.94 \\ 
 
 249.17 & 0.40 &  5.64 &  6.43 &  1.20 &  \ion{Ni}{xvii} &  249.189 & 0.91 \\ 
 
 292.00 & 0.21 &  6.56 &  6.46 &  0.85 &  \ion{Ni}{xviii} &  291.984 & 0.86 \\ 
                                 &   &  &  &  &  \ion{Fe}{xiv} &  292.064 & 0.10 \\ 
 
  93.93 & 0.016 &  6.86 &  6.48 &  0.78 &  \ion{Fe}{xviii} &   93.932 & 0.95 \\

\noalign{\smallskip}\hline 
\end{tabular}
\end{center}
\normalsize
 \label{tab:mh73} 
\end{table}

\section{Summary and conclusions}

Overall, we found a remarkable agreement, to within a relative 20\% accuracy,
 between predicted and observed  quiet Sun line irradiances.
We have considered the spectra 
measured during the extended solar minimum in 2008 by the EVE prototype,
but highlighted several cases where the prototype values appear to 
suffer from calibration problems, by comparing with 
in-flight EVE data and our well-calibrated SoHO CDS irradiances
\citep{delzanna_andretta:2015}.
Comparisons of the PEVE irradiances with those obtained by EVE in 2010
shows  a clear change in all the lines formed above 1 MK, which means that
the only complete EUV spectrum of the quiet Sun available to date is that 
one obtained in 2008, with the few corrections we have noted.

The modeled spectra show a satisfactory level of completeness in the 
 CHIANTI atomic data for these medium-resolution spectra. 
This is very important for modelling purposes.

The good agreement  between predicted and observed  quiet Sun line irradiances
was obtained using the zero-density CHIANTI v.8.07 ion abundances.
Problems with earlier ion charge states for some iron ions were found and fixed.
We experimented with the density-dependent ion abundances calculated 
with the OPEN-ADAS rates, but found several discrepancies, some
by factors of 2. The only improvement  in using the ADAS rates was 
for a few low-temperature ions. 
This clearly indicates the need for improved models of the ion charge states,
now that excitation data within an ion are relatively accurate. 
Work on this issue is in progress. 

The atomic data for \ion{O}{ii} are clearly incorrect, although 
a detailed assessment of such low-temperature lines needs to also 
take into account opacity effects in the lines, which is 
beyond the scope of this paper. 

Contrary to earlier findings  
\citep[see the reviews in][]{delzanna_etal:02_aumic,delzanna_mason:2018},
good agreement between predicted and observed  irradiances is found for 
most anomalous ions  of the Li- and Na-like sequences in the 
upper transition-region and corona. 
Large discrepancies are still present for \ion{O}{vi}, 
some low-temperature ions and the helium lines.
We clearly need improved models to explain the intensities of these lines,
as shown for the helium lines by \cite{golding_etal:2017}.

We confirm our previous findings that the quiet solar corona has photospheric abundances.
Indeed we find excellent agreement in the relative abundances of 
some high-FIP (O, Ne, S) and low-FIP (Fe, Mg, Si, Ca) elements
using  the compilation of photospheric values of \cite{asplund_etal:09}.
We note that the Ne abundance suggested by  \cite{asplund_etal:09}
 was not obtained by direct photospheric measurements. 
The  measurements are sound. They involve strong lines emitted by the whole
solar disk.

The issue of coronal abundances has recently received  renewed interest in 
the solar community. Several  studies based on 
Hinode EIS  have been published \citep[see the review in][]{delzanna_mason:2018}.
 Some of the published results are however puzzling.
For example, several studies  (e.g. \citealt{brooks_etal:2015}
and references therein)  used the  \ion{S}{x} 264.2~\AA\ line 
to measure the sulfur abundance, relative to that of  low-FIP elements
(using \ion{Si}{x}  272.~\AA\ and lines from  iron ions).
Sulfur has a FIP of about 10 eV, but  shows in remote-sensing 
observations  abundance variations  similar 
to those of the high-FIP elements, so it is often used  as a proxy
to measure the FIP bias.
 Note that \ion{S}{x} in the quiet Sun is formed around
1.3 MK (cf. Table~\ref{tab:lines1}).
The results of \cite{brooks_etal:2015} are that the quiet Sun shows an FIP bias of about 2
while active regions have an FIP bias of about 4. 
The quiet Sun results are in contradiction with the present ones,
where the same  \ion{S}{x} 264.2~\AA\  line is well represented (within 20\%) with 
the \cite{asplund_etal:09} photospheric abundances, as shown in Table~\ref{tab:lines1}.
We obtain the same results using the \cite{malinovsky_heroux:73} spectrum.

The active region results of \cite{brooks_etal:2015}  are  in contradiction with 
those obtained  for the 1--2 MK diffuse 
emission in active regions by \citep{delzanna:12_atlas}, 
where an  FIP bias  of about 2 was found.
Clearly, when active regions are present, they produce
all the additional plasma above 1 MK that is not present in the quiet Sun,
as we have discussed in detail in Paper II
\citep{delzanna_andretta:2011} and Paper III \citep{andretta_delzanna:2014}.
The reanalysis of the active Sun observed by \cite{malinovsky_heroux:73}
is consistent with  the quiet solar corona having 
 photospheric abundances, while the 2 MK plasma clearly has an FIP bias of 
about 2.

\begin{acknowledgements} 
Support  by  STFC (UK) via the
consolidated grant of the DAMTP atomic astrophysics group at the University of
Cambridge is acknowledged.  

The calculation of the APAP-Network atomic data used in the present analysis 
was funded by STFC via a grant to the University of Strathclyde.

CHIANTI is a collaborative project involving George Mason University, 
the University of Michigan (USA) and the University of Cambridge (UK).

The irradiance data used here are a courtesy of the NASA/SDO  EVE  science team. 

We acknowledge the use of the OPEN-ADAS database, maintained by the 
 University of Strathclyde.

\end{acknowledgements}


\bibliographystyle{aa}

\bibliography{../bib}  

\begin{thebibliography}{71}
\expandafter\ifx\csname natexlab\endcsname\relax\def\natexlab#1{#1}\fi

\bibitem[{{Andretta} \& {Del Zanna}(2014)}]{andretta_delzanna:2014}
{Andretta}, V. \& {Del Zanna}, G. 2014, \aap, 563, A26

\bibitem[{{Andretta} {et~al.}(2003){Andretta}, {Del Zanna}, \&
  {Jordan}}]{andretta_etal:03}
{Andretta}, V., {Del Zanna}, G., \& {Jordan}, S.~D. 2003, \aap, 400, 737

\bibitem[{{Asplund} {et~al.}(2009){Asplund}, {Grevesse}, {Sauval}, \&
  {Scott}}]{asplund_etal:09}
{Asplund}, M., {Grevesse}, N., {Sauval}, A.~J., \& {Scott}, P. 2009, \araa, 47,
  481

\bibitem[{{Badnell}(2006)}]{badnell:06}
{Badnell}, N.~R. 2006, \apjs, 167, 334

\bibitem[{{Badnell} {et~al.}(2016){Badnell}, {Del Zanna},
  {Fern{\'a}ndez-Menchero}, {Giunta}, {Liang}, {Mason}, \&
  {Storey}}]{badnell_etal:2016}
{Badnell}, N.~R., {Del Zanna}, G., {Fern{\'a}ndez-Menchero}, L., {et~al.} 2016,
  Journal of Physics B Atomic Molecular Physics, 49, 094001

\bibitem[{{Badnell} {et~al.}(2003){Badnell}, {O'Mullane}, {Summers}, {Altun},
  {Bautista}, {Colgan}, {Gorczyca}, {Mitnik}, {Pindzola}, \&
  {Zatsarinny}}]{badnell_etal:03}
{Badnell}, N.~R., {O'Mullane}, M.~G., {Summers}, H.~P., {et~al.} 2003, \aap,
  406, 1151

\bibitem[{{Bradshaw} {et~al.}(2004){Bradshaw}, {Del Zanna}, \&
  {Mason}}]{bradshaw_etal:04}
{Bradshaw}, S.~J., {Del Zanna}, G., \& {Mason}, H.~E. 2004, \aap, 425, 287

\bibitem[{{Brekke} {et~al.}(2000){Brekke}, {Thompson}, {Woods}, \&
  {Eparvier}}]{brekke_etal:00}
{Brekke}, P., {Thompson}, W.~T., {Woods}, T.~N., \& {Eparvier}, F.~G. 2000,
  \apj, 536, 959

\bibitem[{{Brooks} {et~al.}(2015){Brooks}, {Ugarte-Urra}, \&
  {Warren}}]{brooks_etal:2015}
{Brooks}, D.~H., {Ugarte-Urra}, I., \& {Warren}, H.~P. 2015, Nature
  Communications, 6, 5947

\bibitem[{{Burgess} \& {Summers}(1969)}]{burgess_summers:1969}
{Burgess}, A. \& {Summers}, H.~P. 1969, \apj, 157, 1007

\bibitem[{{Burton} {et~al.}(1971){Burton}, {Jordan}, {Ridgeley}, \&
  {Wilson}}]{burton_etal:71}
{Burton}, W.~M., {Jordan}, C., {Ridgeley}, A., \& {Wilson}, R. 1971, Royal
  Society of London Philosophical Transactions Series A, 270, 81

\bibitem[{{Chamberlin} {et~al.}(2007){Chamberlin}, {Hock}, {Crotser},
  {Eparvier}, {Furst}, {Triplett}, {Woodraska}, \&
  {Woods}}]{chamberlin_etal_07}
{Chamberlin}, P.~C., {Hock}, R.~A., {Crotser}, D.~A., {et~al.} 2007, in
  Presented at the Society of Photo-Optical Instrumentation Engineers (SPIE)
  Conference, Vol. 6689, Society of Photo-Optical Instrumentation Engineers
  (SPIE) Conference Series

\bibitem[{{Chamberlin} {et~al.}(2009){Chamberlin}, {Woods}, {Crotser},
  {Eparvier}, {Hock}, \& {Woodraska}}]{chamberlin_etal:09}
{Chamberlin}, P.~C., {Woods}, T.~N., {Crotser}, D.~A., {et~al.} 2009, \grl, 36,
  5102

\bibitem[{{Culhane} {et~al.}(2007){Culhane}, {Harra}, {James}, {Al-Janabi},
  {Bradley}, {Chaudry}, {Rees}, {Tandy}, {Thomas}, {Whillock}, {Winter},
  {Doschek}, {Korendyke}, {Brown}, {Myers}, {Mariska}, {Seely}, {Lang}, {Kent},
  {Shaughnessy}, {Young}, {Simnett}, {Castelli}, {Mahmoud}, {Mapson-Menard},
  {Probyn}, {Thomas}, {Davila}, {Dere}, {Windt}, {Shea}, {Hagood}, {Moye},
  {Hara}, {Watanabe}, {Matsuzaki}, {Kosugi}, {Hansteen}, \&
  {Wikstol}}]{culhane_eis:07}
{Culhane}, J.~L., {Harra}, L.~K., {James}, A.~M., {et~al.} 2007, \solphys, 60

\bibitem[{{Del Zanna}(1999)}]{delzanna_thesis:1999}
{Del Zanna}, G. 1999, PhD thesis, Univ.\ of Central Lancashire, UK

\bibitem[{{Del Zanna}(2009{\natexlab{a}})}]{delzanna:09_fe_7}
{Del Zanna}, G. 2009{\natexlab{a}}, \aap, 508, 501

\bibitem[{{Del Zanna}(2009{\natexlab{b}})}]{delzanna:09_fe_8}
{Del Zanna}, G. 2009{\natexlab{b}}, \aap, 508, 513

\bibitem[{{Del Zanna}(2012{\natexlab{a}})}]{delzanna:12_sxr1}
{Del Zanna}, G. 2012{\natexlab{a}}, \aap, 546, A97

\bibitem[{{Del Zanna}(2012{\natexlab{b}})}]{delzanna:12_atlas}
{Del Zanna}, G. 2012{\natexlab{b}}, \aap, 537, A38

\bibitem[{{Del Zanna}(2013{\natexlab{a}})}]{delzanna:13_eis_calib}
{Del Zanna}, G. 2013{\natexlab{a}}, \aap, 555, A47

\bibitem[{{Del Zanna}(2013{\natexlab{b}})}]{delzanna:2013_multithermal}
{Del Zanna}, G. 2013{\natexlab{b}}, \aap, 558, A73

\bibitem[{{Del Zanna} \& {Andretta}(2011)}]{delzanna_andretta:2011}
{Del Zanna}, G. \& {Andretta}, V. 2011, \aap, 528, A139+

\bibitem[{{Del Zanna} \& {Andretta}(2015)}]{delzanna_andretta:2015}
{Del Zanna}, G. \& {Andretta}, V. 2015, \aap, 584, A29

\bibitem[{{Del Zanna} {et~al.}(2010){Del Zanna}, {Andretta}, {Chamberlin},
  {Woods}, \& {Thompson}}]{delzanna_etal:2010_cdscal}
{Del Zanna}, G., {Andretta}, V., {Chamberlin}, P.~C., {Woods}, T.~N., \&
  {Thompson}, W.~T. 2010, \aap, 518, A49+

\bibitem[{{Del Zanna} {et~al.}(2015{\natexlab{a}}){Del Zanna}, {Andretta},
  {Wieman}, \& {Didkovsky}}]{delzanna_etal:2015_SEM}
{Del Zanna}, G., {Andretta}, V., {Wieman}, S., \& {Didkovsky}, L.
  2015{\natexlab{a}}, \aap, 581, A25

\bibitem[{{Del Zanna} \& {Badnell}(2014)}]{delzanna_badnell:2014_fe_8}
{Del Zanna}, G. \& {Badnell}, N.~R. 2014, \aap, 570, A56

\bibitem[{{Del Zanna} \&
  {Badnell}(2016{\natexlab{a}})}]{delzanna_badnell:2016_s_4}
{Del Zanna}, G. \& {Badnell}, N.~R. 2016{\natexlab{a}}, \mnras, 456, 3720

\bibitem[{{Del Zanna} \&
  {Badnell}(2016{\natexlab{b}})}]{delzanna_badnell:2016_ni_12}
{Del Zanna}, G. \& {Badnell}, N.~R. 2016{\natexlab{b}}, \aap, 585, A118

\bibitem[{{Del Zanna} {et~al.}(2015{\natexlab{b}}){Del Zanna}, {Badnell},
  {Fern{\'a}ndez-Menchero}, {Liang}, {Mason}, \&
  {Storey}}]{delzanna_etal:2015_fe_14}
{Del Zanna}, G., {Badnell}, N.~R., {Fern{\'a}ndez-Menchero}, L., {et~al.}
  2015{\natexlab{b}}, \mnras, 454, 2909

\bibitem[{{Del Zanna} {et~al.}(2004){Del Zanna}, {Berrington}, \&
  {Mason}}]{delzanna_etal:04_fe_10}
{Del Zanna}, G., {Berrington}, K.~A., \& {Mason}, H.~E. 2004, \aap, 422, 731

\bibitem[{{Del Zanna} {et~al.}(2001){Del Zanna}, {Bromage}, {Landi}, \&
  {Landini}}]{delzanna_etal:2001_cdscal}
{Del Zanna}, G., {Bromage}, B.~J.~I., {Landi}, E., \& {Landini}, M. 2001, \aap,
  379, 708

\bibitem[{{Del Zanna} \& {DeLuca}(2018)}]{delzanna_deluca:2018}
{Del Zanna}, G. \& {DeLuca}, E.~E. 2018, \apj, 852, 52

\bibitem[{{Del Zanna} {et~al.}(2015{\natexlab{c}}){Del Zanna}, {Dere}, {Young},
  {Landi}, \& {Mason}}]{delzanna_etal:2015_chianti_v8}
{Del Zanna}, G., {Dere}, K.~P., {Young}, P.~R., {Landi}, E., \& {Mason}, H.~E.
  2015{\natexlab{c}}, \aap, 582, A56

\bibitem[{{Del Zanna} {et~al.}(2002){Del Zanna}, {Landini}, \&
  {Mason}}]{delzanna_etal:02_aumic}
{Del Zanna}, G., {Landini}, M., \& {Mason}, H.~E. 2002, \aap, 385, 968

\bibitem[{{Del Zanna} \& {Mason}(2014)}]{delzanna_mason:2014}
{Del Zanna}, G. \& {Mason}, H.~E. 2014, \aap, 565, A14

\bibitem[{{Del Zanna} \& {Mason}(2018)}]{delzanna_mason:2018}
{Del Zanna}, G. \& {Mason}, H.~E. 2018, Living Reviews in Solar Physics, 15

\bibitem[{{Del Zanna} {et~al.}(2011){Del Zanna}, {O'Dwyer}, \&
  {Mason}}]{delzanna_etal:2011_aia}
{Del Zanna}, G., {O'Dwyer}, B., \& {Mason}, H.~E. 2011, \aap, 535, A46

\bibitem[{{Del Zanna} {et~al.}(2018){Del Zanna}, {Raymond}, {Andretta},
  {Telloni}, \& {Golub}}]{delzanna_etal:2018_cosie}
{Del Zanna}, G., {Raymond}, J., {Andretta}, V., {Telloni}, D., \& {Golub}, L.
  2018, ArXiv e-prints

\bibitem[{{Del Zanna} \& {Storey}(2012)}]{delzanna_storey:12_fe_13}
{Del Zanna}, G. \& {Storey}, P.~J. 2012, \aap, 543, A144

\bibitem[{{Del Zanna} \& {Storey}(2013)}]{delzanna_storey:2013_fe_11}
{Del Zanna}, G. \& {Storey}, P.~J. 2013, \aap, 549, A42

\bibitem[{{Del Zanna} {et~al.}(2012{\natexlab{a}}){Del Zanna}, {Storey},
  {Badnell}, \& {Mason}}]{delzanna_etal:12_fe_10}
{Del Zanna}, G., {Storey}, P.~J., {Badnell}, N.~R., \& {Mason}, H.~E.
  2012{\natexlab{a}}, \aap, 541, A90

\bibitem[{{Del Zanna} {et~al.}(2012{\natexlab{b}}){Del Zanna}, {Storey},
  {Badnell}, \& {Mason}}]{delzanna_etal:12_fe_12}
{Del Zanna}, G., {Storey}, P.~J., {Badnell}, N.~R., \& {Mason}, H.~E.
  2012{\natexlab{b}}, \aap, 543, A139

\bibitem[{{Del Zanna} {et~al.}(2014){Del Zanna}, {Storey}, {Badnell}, \&
  {Mason}}]{delzanna_etal:2014_fe_9}
{Del Zanna}, G., {Storey}, P.~J., {Badnell}, N.~R., \& {Mason}, H.~E. 2014,
  \aap, 565, A77

\bibitem[{{Dere}(2007)}]{dere:07}
{Dere}, K.~P. 2007, \aap, 466, 771

\bibitem[{{Dere} {et~al.}(2019){Dere}, {Del Zanna}, {Young}, {Landi}, \&
  {Sutherland}}]{dere_etal:2019}
{Dere}, K.~P., {Del Zanna}, G., {Young}, P.~R., {Landi}, E., \& {Sutherland},
  R. 2019, \apj, submitted

\bibitem[{{Dere} {et~al.}(1997){Dere}, {Landi}, {Mason}, {Monsignori Fossi}, \&
  {Young}}]{dere_etal:97}
{Dere}, K.~P., {Landi}, E., {Mason}, H.~E., {Monsignori Fossi}, B.~C., \&
  {Young}, P.~R. 1997, \aaps, 125, 149

\bibitem[{{Dere} {et~al.}(2009){Dere}, {Landi}, {Young}, {Del Zanna},
  {Landini}, \& {Mason}}]{dere_etal:09_chianti_v6}
{Dere}, K.~P., {Landi}, E., {Young}, P.~R., {et~al.} 2009, \aap, 498, 915

\bibitem[{{Doschek} {et~al.}(1999){Doschek}, {Laming}, {Doschek}, {Feldman}, \&
  {Wilhelm}}]{doschek_etal:1999}
{Doschek}, E.~E., {Laming}, J.~M., {Doschek}, G.~A., {Feldman}, U., \&
  {Wilhelm}, K. 1999, \apj, 518, 909

\bibitem[{{Dud{\'{\i}}k} {et~al.}(2014){Dud{\'{\i}}k}, {Del Zanna}, {Dzif{\v
  c}{\'a}kov{\'a}}, {Mason}, \& {Golub}}]{dudik_etal:2014_o_4}
{Dud{\'{\i}}k}, J., {Del Zanna}, G., {Dzif{\v c}{\'a}kov{\'a}}, E., {Mason},
  H.~E., \& {Golub}, L. 2014, \apjl, 780, L12

\bibitem[{{Dufresne} \& {Del Zanna}(2018)}]{dufresne_delzanna:2018}
{Dufresne}, R.~P. \& {Del Zanna}, G. 2018, \aap, submitted

\bibitem[{{Fern{\'a}ndez-Menchero}
  {et~al.}(2014{\natexlab{a}}){Fern{\'a}ndez-Menchero}, {Del Zanna}, \&
  {Badnell}}]{fernandez-menchero_etal:2015_be-like}
{Fern{\'a}ndez-Menchero}, L., {Del Zanna}, G., \& {Badnell}, N.~R.
  2014{\natexlab{a}}, \aap, 566, A104

\bibitem[{{Fern{\'a}ndez-Menchero}
  {et~al.}(2014{\natexlab{b}}){Fern{\'a}ndez-Menchero}, {Del Zanna}, \&
  {Badnell}}]{fernandez-menchero_etal:2014_mg-like}
{Fern{\'a}ndez-Menchero}, L., {Del Zanna}, G., \& {Badnell}, N.~R.
  2014{\natexlab{b}}, \aap, 572, A115

\bibitem[{{Fontenla} {et~al.}(1993){Fontenla}, {Avrett}, \&
  {Loeser}}]{fontenla_etal:1993}
{Fontenla}, J.~M., {Avrett}, E.~H., \& {Loeser}, R. 1993, \apj, 406, 319

\bibitem[{{Golding} {et~al.}(2017){Golding}, {Leenaarts}, \&
  {Carlsson}}]{golding_etal:2017}
{Golding}, T.~P., {Leenaarts}, J., \& {Carlsson}, M. 2017, \aap, 597, A102

\bibitem[{{Harrison} {et~al.}(1995){Harrison}, {Sawyer}, {Carter}, {Cruise},
  {Cutler}, {Fludra}, {Hayes}, {Kent}, {Lang}, {Parker}, {Payne}, {Pike},
  {Peskett}, {Richards}, {Gulhane}, {Norman}, {Breeveld}, {Breeveld}, {Al
  Janabi}, {McCalden}, {Parkinson}, {Self}, {Thomas}, {Poland}, {Thomas},
  {Thompson}, {Kjeldseth-Moe}, {Brekke}, {Karud}, {Maltby}, {Aschenbach},
  {Br{\"a}uninger}, {K{\"u}hne}, {Hollandt}, {Siegmund}, {Huber}, {Gabriel},
  {Mason}, \& {Bromage}}]{Harrison-etal:95}
{Harrison}, R.~A., {Sawyer}, E.~C., {Carter}, M.~K., {et~al.} 1995, \solphys,
  162, 233

\bibitem[{{Heroux} {et~al.}(1974){Heroux}, {Cohen}, \&
  {Higgins}}]{heroux_etal:1974}
{Heroux}, L., {Cohen}, M., \& {Higgins}, J.~E. 1974, \jgr, 79, 5237

\bibitem[{{Hock} {et~al.}(2010){Hock}, {Woods}, {Eparvier}, \&
  {Chamberlin}}]{hock_etal:2010}
{Hock}, R., {Woods}, T., {Eparvier}, F., \& {Chamberlin}, P. 2010, in COSPAR
  Meeting, Vol.~38, 38th COSPAR Scientific Assembly, 5

\bibitem[{{Laming}(2015)}]{laming:2015}
{Laming}, J.~M. 2015, Living Reviews in Solar Physics, 12

\bibitem[{{Laming} {et~al.}(1995){Laming}, {Drake}, \&
  {Widing}}]{laming_etal:1995}
{Laming}, J.~M., {Drake}, J.~J., \& {Widing}, K.~G. 1995, \apj, 443, 416

\bibitem[{{Landi} {et~al.}(2002){Landi}, {Feldman}, \&
  {Dere}}]{landi_etal:2002_cds}
{Landi}, E., {Feldman}, U., \& {Dere}, K.~P. 2002, \apj, 574, 495

\bibitem[{Landi {et~al.}(2002)Landi, Feldman, \& Dere}]{landi_etal:2002}
Landi, E., Feldman, U., \& Dere, K.~P. 2002, \apjs, 139, 281

\bibitem[{{Landi} \& {Young}(2009)}]{landi_young:2009}
{Landi}, E. \& {Young}, P.~R. 2009, \apj, 706, 1

\bibitem[{{Landi} {et~al.}(2013){Landi}, {Young}, {Dere}, {Del Zanna}, \&
  {Mason}}]{landi_etal:12_chianti_v7.1}
{Landi}, E., {Young}, P.~R., {Dere}, K.~P., {Del Zanna}, G., \& {Mason}, H.~E.
  2013, \apj, 763, 86

\bibitem[{{Malinovsky} \& {Heroux}(1973)}]{malinovsky_heroux:73}
{Malinovsky}, L. \& {Heroux}, M. 1973, \apj, 181, 1009

\bibitem[{{Thomas} \& {Neupert}(1994)}]{thomas_neupert:94}
{Thomas}, R.~J. \& {Neupert}, W.~M. 1994, \apjs, 91, 461

\bibitem[{{Wang} {et~al.}(2011){Wang}, {Thomas}, {Brosius}, {Young}, {Rabin},
  {Davila}, \& {Del Zanna}}]{wang_etal:2011}
{Wang}, T., {Thomas}, R.~J., {Brosius}, J.~W., {et~al.} 2011, \apjs, 197, 32

\bibitem[{Wilhelm {et~al.}(1995)}]{Wilhelm-etal:95}
Wilhelm, K. {et~al.} 1995, 162, 189

\bibitem[{{Woods} {et~al.}(2009){Woods}, {Chamberlin}, {Harder}, {Hock},
  {Snow}, {Eparvier}, {Fontenla}, {McClintock}, \& {Richard}}]{woods_etal:09}
{Woods}, T.~N., {Chamberlin}, P.~C., {Harder}, J.~W., {et~al.} 2009, \grl, 36,
  1101

\bibitem[{{Woods} {et~al.}(2005){Woods}, {Eparvier}, {Bailey}, {Chamberlin},
  {Lean}, {Rottman}, {Solomon}, {Tobiska}, \& {Woodraska}}]{woods_etal:05}
{Woods}, T.~N., {Eparvier}, F.~G., {Bailey}, S.~M., {et~al.} 2005, Journal of
  Geophysical Research (Space Physics), 110, 1312

\bibitem[{{Woods} {et~al.}(2012){Woods}, {Eparvier}, {Hock}, {Jones},
  {Woodraska}, {Judge}, {Didkovsky}, {Lean}, {Mariska}, {Warren}, {McMullin},
  {Chamberlin}, {Berthiaume}, {Bailey}, {Fuller-Rowell}, {Sojka}, {Tobiska}, \&
  {Viereck}}]{woods_etal:12}
{Woods}, T.~N., {Eparvier}, F.~G., {Hock}, R., {et~al.} 2012, \solphys, 275,
  115

\bibitem[{{Young} {et~al.}(1998){Young}, {Landi}, \& {Thomas}}]{young_etal:98}
{Young}, P.~R., {Landi}, E., \& {Thomas}, R.~J. 1998, \aap, 329, 291

\end{thebibliography}

\appendix

\section{Line list}

Table~\ref{tab:lines2} presents the full list of observed and predicted EUV irradiances 
for the main, stronger lines in the 60--1040~\AA\ range.

\begin{longtable}[c]{@{}rrcclrrl@{}}
\caption{EUV irradiances of the quiet Sun. Same layout as in 
Table~\ref{tab:lines1}. } \\
 \label{tab:lines2} \\
\hline
$\lambda_{\rm obs}$(\AA)  & $I_{\rm obs}$  & log $T$ [K]  & $R$ & Ion & $\lambda_{\rm exp}$(\AA) &  ratio \\
   \hline
\endhead

\hline \multicolumn{3}{|r|}{{Continued on next page}} \\ \hline
\endfoot

  86.91 & 0.18 &  5.93 &  6.04 &  0.39 &  \ion{Mg}{viii} &   87.022 & 0.14 \\ 
                                 &   &  &  &  &  \ion{Mg}{viii} &   86.844 & 0.21 \\ 
                                 &   &  &  &  &  \ion{Si}{vii} &   86.914 & 0.13 \\ 
                                 &   &  &  &  &  \ion{Fe}{xi} &   86.772 & 0.31 \\ 
 
  88.13 & 0.07 &  5.84 &  6.03 &  0.77 &  \ion{Ne}{viii} &   88.120 & 0.25 \\ 
                                 &   &  &  &  &  \ion{Ne}{viii} &   88.092 & 0.49 \\ 
 
  88.98 & 0.08 &  6.15 &  6.11 &  0.90 &  \ion{Fe}{xi} &   89.178 & 0.19 \\ 
                                 &   &  &  &  &  \ion{Fe}{xi} &   88.933 & 0.63 \\

  94.14 & 0.12 &  6.04 &  6.05 &  0.40 &  \ion{Fe}{x} &   94.012 & 0.74 \\ 
 
  95.42 & 0.10 &  5.74 &  5.95 &  0.54 &  \ion{Mg}{vii} &   95.423 & 0.12 \\ 
                                 &   &  &  &  &  \ion{Mg}{vi} &   95.483 & 0.12 \\ 
                                 &   &  &  &  &  \ion{Si}{vi} &   95.555 & 0.12 \\ 
                                 &   &  &  &  &  \ion{Fe}{x} &   95.339 & 0.10 \\ 
                                 &   &  &  &  &  \ion{Fe}{x} &   95.374 & 0.20 \\ 
 
  96.10 & 0.16 &  6.06 &  6.05 &  0.75 &  \ion{Fe}{x} &   96.121 & 0.31 \\ 
                                 &   &  &  &  &  \ion{Fe}{x} &   96.007 & 0.59 \\

  98.12 & 0.22 &  5.80 &  6.01 &  0.54 &  \ion{Ne}{viii} &   98.116 & 0.20 \\ 
                                 &   &  &  &  &  \ion{Ne}{viii} &   98.261 & 0.35 \\ 
                                 &   &  &  &  &  \ion{Ne}{vii} &   97.502 & 0.11 \\ 
                                 &   &  &  &  &  \ion{Fe}{x} &   97.838 & 0.14 \\

 100.67 & 0.10 &  5.70 &  6.04 &  0.54 &  \ion{Si}{vi} &  100.953 & 0.13 \\ 
                                 &   &  &  &  &  \ion{Fe}{xi} &  100.575 & 0.59 \\

 103.40 & 0.16 &  5.91 &  6.02 &  0.86 &  \ion{Ne}{viii} &  103.085 & 0.19 \\ 
                                 &   &  &  &  &  \ion{Fe}{ix} &  103.566 & 0.67 \\ 
 
 105.20 & 0.12 &  5.91 &  6.01 &  0.48 &  \ion{Fe}{ix} &  105.208 & 0.87 \\

 111.59 & 0.17 &  5.72 &  5.95 &  0.34 &  \ion{Mg}{vii} &  111.984 & 0.12 \\ 
                                 &   &  &  &  &  \ion{Fe}{ix} &  111.791 & 0.11 \\ 
                                 &   &  &  &  &  \ion{Fe}{ix} &  112.096 & 0.19 \\ 
 
 113.75 & 0.10 &  5.93 &  6.01 &  1.12 &  \ion{Fe}{ix} &  113.793 & 0.46 \\ 
                                 &   &  &  &  &  \ion{Fe}{ix} &  114.024 & 0.24 \\ 
                                 &   &  &  &  &  \ion{Fe}{ix} &  114.111 & 0.11 \\ 
  
 116.87 & 0.13 &  5.75 &  5.99 &  0.66 &  \ion{Ne}{vii} &  116.691 & 0.24 \\ 
                                 &   &  &  &  &  \ion{Fe}{ix} &  116.803 & 0.14 \\ 
                                 &   &  &  &  &  \ion{Fe}{ix} &  116.803 & 0.15 \\

 131.12 & 0.11 &  5.74 &  5.93 &  0.96 &  \ion{Fe}{viii} &  130.941 & 0.37 \\ 
                                 &   &  &  &  &  \ion{Fe}{viii} &  131.240 & 0.56 \\

 144.50 & 0.17 &  5.66 &  6.00 &  0.03 &  \ion{Ne}{vi} &  144.881 & 0.14 \\ 
                                 &   &  &  &  &  \ion{Fe}{x} &  144.328 & 0.59 \\ 
 
 145.73 & 0.12 &  5.56 &  5.79 &  0.03 &  \ion{Mg}{v} &  146.083 & 0.52 \\ 
                                 &   &  &  &  &  \ion{Mg}{v} &  145.488 & 0.12 \\ 
                                 &   &  &  &  &  \ion{Si}{vi} &  145.342 & 0.11 \\ 
                                 &   &  &  &  &  \ion{Fe}{x} &  145.754 & 0.14 \\ 
 
 148.38 & 0.38 &  6.11 &  6.08 &  1.14 &  \ion{Ni}{xi} &  148.377 & 0.99 \\ 
 
 150.12 & 0.19 &  5.52 &  5.96 &  0.42 &  \ion{O}{vi} &  150.125 & 0.33 \\ 
                                 &   &  &  &  &  \ion{O}{vi} &  150.089 & 0.66 \\ 
 
 152.05 & 0.11 &  6.24 &  6.14 &  0.80 &  \ion{Ni}{xii} &  152.151 & 0.94 \\

 167.56 & 0.32 &  5.70 &  5.91 &  0.88 &  \ion{Fe}{viii} &  167.486 & 0.85 \\ 
 
 168.38 & 0.28 &  5.72 &  5.92 &  0.82 &  \ion{Fe}{viii} &  168.544 & 0.98 \\ 
 
 168.69 & 0.17 &  5.70 &  5.92 &  2.28 &  \ion{Fe}{viii} &  168.929 & 0.31 \\ 
                                 &   &  &  &  &  \ion{Fe}{viii} &  168.544 & 0.60 \\ 
 
 170.98 & 7.0 &  5.91 &  6.00 &  0.95 &  \ion{Fe}{ix} &  171.073 & 0.99 \\ 
 
 173.01 & 0.33 &  5.52 &  5.96 &  0.52 &  \ion{O}{vi} &  172.935 & 0.29 \\ 
                                 &   &  &  &  &  \ion{O}{vi} &  173.079 & 0.52 \\ 
                                 &   &  &  &  &  \ion{Fe}{ix} &  173.224 & 0.10 \\ 
 
 174.57 & 3.7 &  6.05 &  6.06 &  1.03 &  \ion{Fe}{x} &  174.531 & 0.98 \\ 
 
 177.20 & 2.3 &  6.04 &  6.05 &  1.11 &  \ion{Fe}{x} &  177.240 & 0.83 \\ 
 
 178.09 & 0.14 &  6.12 &  6.10 &  0.78 &  \ion{Fe}{xi} &  178.058 & 0.91 \\ 
 
 180.41 & 2.7 &  6.13 &  6.10 &  1.07 &  \ion{Fe}{xi} &  180.401 & 0.84 \\ 
 
 182.17 & 0.54 &  6.12 &  6.10 &  0.80 &  \ion{Fe}{xi} &  182.167 & 0.85 \\ 
                                 &   &  &  &  &  \ion{Fe}{x} &  182.307 & 0.11 \\ 
 
 184.46 & 1.2 &  6.05 &  6.05 &  0.83 &  \ion{Fe}{x} &  184.537 & 0.85 \\ 
 
 185.29 & 0.35 &  5.71 &  5.92 &  1.12 &  \ion{Fe}{viii} &  185.213 & 0.88 \\ 
 
 186.80 & 0.57 &  5.71 &  6.06 &  1.07 &  \ion{Fe}{xii} &  186.887 & 0.39 \\ 
                                 &   &  &  &  &  \ion{Fe}{viii} &  186.598 & 0.39 \\ 
 
 188.29 & 2.3 &  6.13 &  6.09 &  1.08 &  \ion{Fe}{xi} &  188.216 & 0.50 \\ 
                                 &   &  &  &  &  \ion{Fe}{xi} &  188.299 & 0.31 \\ 
                                 &   &  &  &  &  \ion{Fe}{ix} &  188.493 & 0.13 \\ 
 
 190.00 & 0.66 &  6.02 &  6.05 &  0.92 &  \ion{Fe}{x} &  190.037 & 0.47 \\ 
                                 &   &  &  &  &  \ion{Fe}{ix} &  189.935 & 0.31 \\ 
 
 192.35 & 0.64 &  6.19 &  6.12 &  0.82 &  \ion{Fe}{xii} &  192.394 & 0.76 \\ 
 
 192.80 & 0.26 &  5.42 &  6.03 &  1.45 &  \ion{O}{v} &  192.904 & 0.11 \\ 
                                 &   &  &  &  &  \ion{Fe}{xi} &  192.813 & 0.70 \\ 
 
 193.51 & 1.0 &  6.19 &  6.14 &  1.00 &  \ion{Fe}{xii} &  193.509 & 0.83 \\ 
                                 &   &  &  &  &  \ion{Fe}{x} &  193.715 & 0.10 \\ 
 
 195.09 & 1.4 &  6.20 &  6.15 &  0.96 &  \ion{Fe}{xii} &  195.119 & 0.95 \\ 
 
 196.10 & 0.19 &  5.31 &  5.89 &  0.56 &  \ion{O}{iv} &  196.006 & 0.18 \\ 
                                 &   &  &  &  &  \ion{Fe}{xiii} &  196.525 & 0.18 \\ 
                                 &   &  &  &  &  \ion{Fe}{viii} &  195.972 & 0.44 \\ 
 
 197.50 & 0.21 &  5.92 &  6.03 &  1.14 &  \ion{Fe}{xiii} &  197.431 & 0.15 \\ 
                                 &   &  &  &  &  \ion{Fe}{ix} &  197.854 & 0.73 \\ 
 
 198.51 & 0.26 &  5.96 &  6.04 &  0.84 &  \ion{S}{viii} &  198.554 & 0.65 \\ 
                                 &   &  &  &  &  \ion{Fe}{xi} &  198.538 & 0.27 \\ 
 
 200.28 & 0.14 &  6.24 &  6.15 &  0.51 &  \ion{Fe}{xiii} &  200.021 & 0.74 \\ 
                                 &   &  &  &  &  \ion{Fe}{xii} &  200.355 & 0.10 \\ 
                                 &   &  &  &  &  \ion{Fe}{ix} &  199.975 & 0.10 \\ 
 
 201.06 & 0.14 &  6.24 &  6.15 &  1.55 &  \ion{Fe}{xiii} &  201.126 & 0.64 \\ 
                                 &   &  &  &  &  \ion{Fe}{xi} &  201.112 & 0.32 \\ 
 
 202.10 & 0.96 &  6.25 &  6.15 &  0.92 &  \ion{Fe}{xiii} &  202.044 & 0.64 \\ 
                                 &   &  &  &  &  \ion{Fe}{xi} &  202.424 & 0.12 \\ 
 
 203.89 & 0.31 &  6.23 &  6.16 &  1.26 &  \ion{Fe}{xiii} &  203.795 & 0.18 \\ 
                                 &   &  &  &  &  \ion{Fe}{xiii} &  203.826 & 0.41 \\ 
                                 &   &  &  &  &  \ion{Fe}{xii} &  203.728 & 0.22 \\ 
 
 206.22 & 0.20 &  6.14 &  6.11 &  0.59 &  \ion{Fe}{xii} &  206.368 & 0.18 \\ 
                                 &   &  &  &  &  \ion{Fe}{xi} &  206.258 & 0.41 \\ 
                                 &   &  &  &  &  \ion{Fe}{xi} &  206.169 & 0.35 \\ 
 
 207.67 & 0.33 &  6.06 &  6.05 &  0.80 &  \ion{Fe}{x} &  207.449 & 0.46 \\ 
                                 &   &  &  &  &  \ion{Ni}{xi} &  207.922 & 0.33 \\ 
 
 209.88 & 0.28 &  6.25 &  6.16 &  0.70 &  \ion{Fe}{xiii} &  209.619 & 0.19 \\ 
                                 &   &  &  &  &  \ion{Fe}{xiii} &  209.916 & 0.52 \\ 
                                 &   &  &  &  &  \ion{Fe}{xi} &  209.771 & 0.22 \\ 
 
 211.40 & 0.33 &  6.29 &  6.19 &  0.97 &  \ion{Fe}{xiv} &  211.317 & 0.75 \\ 
                                 &   &  &  &  &  \ion{Ni}{xi} &  211.430 & 0.12 \\ 
 
 215.00 & 0.24 &  5.42 &  5.84 &  0.58 &  \ion{O}{v} &  215.103 & 0.15 \\ 
                                 &   &  &  &  &  \ion{O}{v} &  215.245 & 0.25 \\ 
                                 &   &  &  &  &  \ion{Si}{viii} &  214.759 & 0.37 \\ 
 
 217.04 & 0.69 &  5.90 &  6.01 &  0.81 &  \ion{Si}{viii} &  216.922 & 0.26 \\ 
                                 &   &  &  &  &  \ion{Fe}{ix} &  217.101 & 0.58 \\ 
 
 218.85 & 0.35 &  5.88 &  6.04 &  0.69 &  \ion{Fe}{xiv} &  219.130 & 0.13 \\ 
                                 &   &  &  &  &  \ion{Fe}{ix} &  218.937 & 0.78 \\ 
 
 220.21 & 0.46 &  5.43 &  6.04 &  0.73 &  \ion{O}{v} &  220.353 & 0.14 \\ 
                                 &   &  &  &  &  \ion{Fe}{xiv} &  220.085 & 0.16 \\ 
                                 &   &  &  &  &  \ion{Fe}{x} &  220.247 & 0.54 \\ 
 
 221.23 & 0.23 &  6.22 &  6.11 &  0.66 &  \ion{S}{ix} &  221.241 & 0.49 \\ 
                                 &   &  &  &  &  \ion{Fe}{xii} &  220.870 & 0.26 \\ 
 
 222.03 & 0.23 &  6.23 &  6.08 &  0.51 &  \ion{Fe}{xiii} &  221.828 & 0.39 \\ 
                                 &   &  &  &  &  \ion{Fe}{x} &  221.689 & 0.15 \\ 
                                 &   &  &  &  &  \ion{Fe}{viii} &  222.190 & 0.29 \\ 
 
 223.03 & 0.24 &  6.21 &  6.11 &  0.22 &  \ion{S}{ix} &  223.262 & 0.16 \\ 
                                 &   &  &  &  &  \ion{Fe}{xii} &  222.964 & 0.32 \\ 
                                 &   &  &  &  &  \ion{Fe}{xii} &  223.000 & 0.34 \\ 
 
 223.86 & 0.35 &  5.69 &  6.03 &  0.57 &  \ion{Si}{ix} &  223.744 & 0.47 \\ 
                                 &   &  &  &  &  \ion{Fe}{viii} &  224.305 & 0.28 \\ 
 
 224.95 & 1.1 &  6.06 &  6.06 &  0.65 &  \ion{Si}{ix} &  225.025 & 0.41 \\ 
                                 &   &  &  &  &  \ion{S}{ix} &  224.726 & 0.46 \\ 
 
 226.13 & 0.49 &  6.04 &  6.06 &  0.49 &  \ion{Fe}{x} &  225.856 & 0.95 \\ 
 
 227.00 & 1.0 &  6.05 &  6.06 &  0.78 &  \ion{Si}{ix} &  227.002 & 0.60 \\ 
                                 &   &  &  &  &  \ion{Fe}{x} &  226.998 & 0.18 \\ 
 
 228.21 & 0.38 &  6.24 &  6.15 &  0.11 &  \ion{S}{x} &  228.166 & 0.10 \\ 
                                 &   &  &  &  &  \ion{Fe}{xiii} &  228.160 & 0.76 \\ 
 
 229.28 & 0.26 &  6.06 &  6.07 &  0.42 &  \ion{S}{ix} &  228.832 & 0.94 \\

 233.41 & 0.39 &  5.30 &  6.01 &  0.24 &  \ion{Si}{viii} &  233.139 & 0.15 \\ 
                                 &   &  &  &  &  \ion{Fe}{x} &  233.445 & 0.61 \\ 
 
 234.38 & 0.60 &  6.06 &  6.06 &  0.39 &  \ion{Fe}{xi} &  234.730 & 0.24 \\ 
                                 &   &  &  &  &  \ion{Fe}{x} &  234.315 & 0.67 \\ 
 
 237.16 & 0.46 &  4.95 &  5.85 &  0.23 &  \ion{He}{ii} &  237.331 & 0.16 \\ 
                                 &   &  &  &  &  \ion{He}{ii} &  237.331 & 0.32 \\ 
                                 &   &  &  &  &  \ion{Fe}{xi} &  237.262 & 0.22 \\ 
                                 &   &  &  &  &  \ion{Fe}{viii} &  237.291 & 0.18 \\ 
 
 238.56 & 0.42 &  5.29 &  5.42 &  0.65 &  \ion{O}{iv} &  238.360 & 0.29 \\ 
                                 &   &  &  &  &  \ion{O}{iv} &  238.570 & 0.53 \\ 
                                 &   &  &  &  &  \ion{Fe}{viii} &  238.329 & 0.10 \\ 
 
 240.57 & 0.34 &  6.18 &  6.12 &  0.89 &  \ion{Fe}{xiii} &  240.696 & 0.14 \\ 
                                 &   &  &  &  &  \ion{Fe}{xii} &  240.740 & 0.15 \\ 
                                 &   &  &  &  &  \ion{Fe}{xi} &  240.717 & 0.64 \\ 
 
 241.73 & 1.2 &  5.89 &  5.99 &  0.99 &  \ion{Fe}{ix} &  241.739 & 0.98 \\

 242.95 & 0.89 &  4.95 &  5.80 &  0.20 &  \ion{He}{ii} &  243.027 & 0.24 \\ 
                                 &   &  &  &  &  \ion{He}{ii} &  243.027 & 0.47 \\ 
                                 &   &  &  &  &  \ion{S}{xi} &  242.850 & 0.12 \\

 244.84 & 0.62 &  5.87 &  5.98 &  0.90 &  \ion{Fe}{ix} &  244.909 & 0.98 \\ 
 
 246.15 & 0.21 &  5.64 &  6.08 &  0.88 &  \ion{Si}{vi} &  246.003 & 0.31 \\ 
                                 &   &  &  &  &  \ion{Fe}{xiii} &  246.209 & 0.58 \\ 
 
 247.30 & 0.25 &  5.28 &  6.07 &  0.40 &  \ion{S}{xi} &  247.159 & 0.22 \\ 
                                 &   &  &  &  &  \ion{S}{xi} &  246.895 & 0.15 \\ 
                                 &   &  &  &  &  \ion{Fe}{xi} &  247.291 & 0.27 \\ 
 
 248.45 & 0.19 &  5.42 &  5.80 &  0.66 &  \ion{O}{v} &  248.460 & 0.59 \\ 
                                 &   &  &  &  &  \ion{Al}{viii} &  248.459 & 0.18 \\ 
                                 &   &  &  &  &  \ion{Fe}{xii} &  248.500 & 0.17 \\

 251.98 & 0.24 &  6.25 &  6.18 &  1.15 &  \ion{Fe}{xiii} &  251.952 & 0.74 \\ 
                                 &   &  &  &  &  \ion{Fe}{xii} &  251.868 & 0.13 \\ 
 
 253.87 & 0.26 &  5.67 &  6.05 &  0.67 &  \ion{Si}{x} &  253.790 & 0.65 \\ 
                                 &   &  &  &  &  \ion{Fe}{viii} &  253.956 & 0.32 \\

 256.28 & 2.5 &  4.94 &  5.89 &  0.46 &  \ion{He}{ii} &  256.318 & 0.16 \\ 
                                 &   &  &  &  &  \ion{He}{ii} &  256.317 & 0.32 \\ 
                                 &   &  &  &  &  \ion{Si}{x} &  256.377 & 0.29 \\ 
                                 &   &  &  &  &  \ion{Fe}{x} &  256.398 & 0.10 \\ 
 
 257.23 & 2.2 &  6.08 &  6.08 &  0.79 &  \ion{Fe}{xi} &  256.919 & 0.14 \\ 
                                 &   &  &  &  &  \ion{Fe}{x} &  257.259 & 0.16 \\ 
                                 &   &  &  &  &  \ion{Fe}{x} &  257.263 & 0.44 \\ 
 
 258.32 & 0.74 &  6.14 &  6.11 &  0.87 &  \ion{Si}{x} &  258.374 & 0.91 \\ 
 
 259.55 & 0.31 &  6.18 &  6.13 &  0.83 &  \ion{S}{x} &  259.497 & 0.70 \\ 
                                 &   &  &  &  &  \ion{Fe}{xii} &  259.418 & 0.13 \\ 
                                 &   &  &  &  &  \ion{Fe}{xii} &  259.973 & 0.11 \\ 
 
 261.06 & 0.41 &  6.15 &  6.11 &  0.74 &  \ion{Si}{x} &  261.056 & 0.96 \\ 
 
 263.05 & 0.022 &  6.43 &  6.26 &  0.67 &  \ion{Fe}{xvi} &  262.976 & 0.45 \\ 
                                 &   &  &  &  &  \ion{Fe}{xiii} &  262.984 & 0.28 \\ 
                                 &   &  &  &  &  \ion{Mn}{viii} &  263.163 & 0.15 \\ 
 
 264.30 & 0.38 &  6.18 &  6.13 &  0.71 &  \ion{S}{x} &  264.231 & 0.98 \\

 270.47 & 0.10 &  6.29 &  6.09 &  1.26 &  \ion{Mg}{vi} &  270.391 & 0.31 \\ 
                                 &   &  &  &  &  \ion{Fe}{xiv} &  270.521 & 0.59 \\ 
 
 272.05 & 0.42 &  5.28 &  6.03 &  0.79 &  \ion{Si}{x} &  271.992 & 0.75 \\ 
 
 274.19 & 0.16 &  6.29 &  6.17 &  1.43 &  \ion{Si}{vii} &  274.180 & 0.22 \\ 
                                 &   &  &  &  &  \ion{Fe}{xiv} &  274.204 & 0.74 \\ 
 
 275.42 & 0.35 &  5.80 &  5.93 &  1.08 &  \ion{Si}{vii} &  275.361 & 0.84 \\ 
                                 &   &  &  &  &  \ion{Si}{vii} &  275.676 & 0.14 \\ 
 
 277.03 & 0.79 &  5.84 &  6.02 &  0.90 &  \ion{Si}{x} &  277.264 & 0.27 \\ 
                                 &   &  &  &  &  \ion{Mg}{vii} &  277.003 & 0.18 \\ 
                                 &   &  &  &  &  \ion{Si}{viii} &  276.850 & 0.15 \\ 
                                 &   &  &  &  &  \ion{Si}{viii} &  277.058 & 0.25 \\ 
                                 &   &  &  &  &  \ion{Si}{vii} &  276.851 & 0.10 \\ 
 
 278.40 & 0.33 &  5.80 &  5.92 &  1.05 &  \ion{Mg}{vii} &  278.404 & 0.63 \\ 
                                 &   &  &  &  &  \ion{Si}{vii} &  278.450 & 0.30 \\ 
 
 279.90 & 0.15 &  5.26 &  5.50 &  1.07 &  \ion{O}{iv} &  279.631 & 0.28 \\ 
                                 &   &  &  &  &  \ion{O}{iv} &  279.933 & 0.56 \\ 
                                 &   &  &  &  &  \ion{Al}{ix} &  280.151 & 0.12 \\

 284.12 & 0.46 &  6.34 &  6.24 &  1.05 &  \ion{Al}{ix} &  284.042 & 0.20 \\ 
                                 &   &  &  &  &  \ion{Fe}{xv} &  284.163 & 0.76 \\ 
 
 290.78 & 0.32 &  6.14 &  6.08 &  0.76 &  \ion{Si}{ix} &  290.687 & 0.61 \\ 
                                 &   &  &  &  &  \ion{Fe}{xii} &  291.010 & 0.32 \\ 
 
 292.83 & 0.53 &  6.05 &  6.06 &  0.90 &  \ion{Si}{ix} &  292.809 & 0.37 \\ 
                                 &   &  &  &  &  \ion{Si}{ix} &  292.855 & 0.30 \\ 
                                 &   &  &  &  &  \ion{Si}{ix} &  292.759 & 0.27 \\ 
 
 296.15 & 0.76 &  6.05 &  6.06 &  0.93 &  \ion{Si}{ix} &  296.211 & 0.26 \\ 
                                 &   &  &  &  &  \ion{Si}{ix} &  296.113 & 0.72 \\

 303.76 & 63.1 &  4.92 &  5.24 &  0.10 &  \ion{He}{ii} &  303.786 & 0.33 \\ 
                                 &   &  &  &  &  \ion{He}{ii} &  303.781 & 0.65 \\ 
 
 304.71 & 0.64 &  5.25 &  6.04 &  0.05 &  \ion{Al}{ix} &  305.045 & 0.82 \\

 313.77 & 0.56 &  5.91 &  5.97 &  0.83 &  \ion{Mg}{viii} &  313.743 & 0.93 \\ 
 
 314.57 & 0.60 &  5.94 &  6.00 &  0.70 &  \ion{Si}{viii} &  314.356 & 0.94 \\ 
 
 315.07 & 1.1 &  5.91 &  5.98 &  1.03 &  \ion{Mg}{viii} &  315.015 & 0.97 \\ 
 
 316.22 & 0.86 &  5.94 &  6.01 &  0.95 &  \ion{Si}{viii} &  316.218 & 0.97 \\ 
 
 317.05 & 0.44 &  5.91 &  5.98 &  0.82 &  \ion{Mg}{viii} &  317.028 & 0.78 \\ 
                                 &   &  &  &  &  \ion{Fe}{ix} &  317.193 & 0.19 \\

 319.86 & 1.4 &  5.94 &  6.01 &  0.89 &  \ion{Si}{viii} &  319.840 & 0.98 \\

 328.29 & 0.13 &  5.08 &  6.02 &  1.13 &  \ion{O}{iii} &  328.448 & 0.15 \\ 
                                 &   &  &  &  &  \ion{Al}{viii} &  328.184 & 0.31 \\ 
                                 &   &  &  &  &  \ion{Cr}{xiii} &  328.268 & 0.42 \\ 
 
 332.82 & 0.37 &  6.11 &  6.10 &  0.76 &  \ion{Al}{x} &  332.790 & 0.97 \\

 335.31 & 0.48 &  6.42 &  6.13 &  0.92 &  \ion{Mg}{viii} &  335.231 & 0.51 \\ 
                                 &   &  &  &  &  \ion{Fe}{xvi} &  335.409 & 0.23 \\ 
                                 &   &  &  &  &  \ion{Fe}{xii} &  335.380 & 0.11 \\ 
                                 &   &  &  &  &  \ion{Fe}{ix} &  335.290 & 0.11 \\

 339.05 & 0.32 &  5.91 &  5.98 &  0.91 &  \ion{Mg}{viii} &  338.983 & 0.96 \\ 
 
 341.18 & 0.27 &  6.09 &  6.05 &  1.03 &  \ion{Fe}{xi} &  341.113 & 0.55 \\ 
                                 &   &  &  &  &  \ion{Fe}{ix} &  341.396 & 0.11 \\ 
                                 &   &  &  &  &  \ion{Fe}{ix} &  341.159 & 0.29 \\ 
 
 342.10 & 0.90 &  5.68 &  6.04 &  0.31 &  \ion{Si}{ix} &  341.951 & 0.84 \\ 
 
 345.13 & 0.90 &  6.06 &  6.06 &  0.72 &  \ion{Si}{ix} &  345.121 & 0.74 \\ 
                                 &   &  &  &  &  \ion{Si}{ix} &  344.954 & 0.21 \\ 
 
 345.75 & 0.65 &  6.04 &  6.05 &  0.85 &  \ion{Fe}{x} &  345.738 & 0.98 \\ 
 
 347.45 & 1.2 &  6.15 &  6.11 &  0.61 &  \ion{Si}{x} &  347.402 & 0.95 \\

 349.10 & 0.56 &  5.65 &  5.90 &  0.29 &  \ion{Mg}{vi} &  349.125 & 0.25 \\ 
                                 &   &  &  &  &  \ion{Mg}{vi} &  349.164 & 0.34 \\ 
                                 &   &  &  &  &  \ion{Fe}{xi} &  349.046 & 0.23 \\ 
 
 349.91 & 0.96 &  6.05 &  6.06 &  0.67 &  \ion{Si}{ix} &  349.792 & 0.14 \\ 
                                 &   &  &  &  &  \ion{Si}{ix} &  349.860 & 0.83 \\ 
 
 352.57 & 1.1 &  6.12 &  6.08 &  0.61 &  \ion{Fe}{xi} &  352.670 & 0.91 \\ 
 
 354.40 & 0.33 &  5.51 &  5.72 &  0.15 &  \ion{Mg}{v} &  354.221 & 0.40 \\ 
                                 &   &  &  &  &  \ion{Ca}{viii} &  354.167 & 0.43 \\ 
                                 &   &  &  &  &  \ion{Ca}{vii} &  354.419 & 0.12 \\ 
 
 356.10 & 0.64 &  6.14 &  6.10 &  0.75 &  \ion{Si}{x} &  356.049 & 0.20 \\ 
                                 &   &  &  &  &  \ion{Si}{x} &  356.037 & 0.75 \\

 358.63 & 0.31 &  5.37 &  5.84 &  0.85 &  \ion{Ne}{v} &  358.476 & 0.25 \\ 
                                 &   &  &  &  &  \ion{Ne}{iv} &  358.694 & 0.27 \\ 
                                 &   &  &  &  &  \ion{Fe}{xi} &  358.613 & 0.38 \\ 
 
 359.55 & 0.26 &  5.46 &  5.99 &  1.25 &  \ion{Ne}{v} &  359.375 & 0.34 \\ 
                                 &   &  &  &  &  \ion{Ca}{viii} &  359.367 & 0.11 \\ 
                                 &   &  &  &  &  \ion{Fe}{xiii} &  359.839 & 0.15 \\ 
                                 &   &  &  &  &  \ion{Fe}{xiii} &  359.644 & 0.35 \\

 364.54 & 0.67 &  6.19 &  6.14 &  0.83 &  \ion{Fe}{xii} &  364.467 & 0.92 \\ 
 
 365.50 & 0.57 &  5.45 &  5.94 &  1.01 &  \ion{Ne}{v} &  365.603 & 0.17 \\ 
                                 &   &  &  &  &  \ion{Mg}{vii} &  365.181 & 0.13 \\ 
                                 &   &  &  &  &  \ion{Mg}{vii} &  365.247 & 0.10 \\ 
                                 &   &  &  &  &  \ion{Mg}{vii} &  365.238 & 0.14 \\ 
                                 &   &  &  &  &  \ion{Fe}{x} &  365.560 & 0.41 \\ 
 
 368.10 & 5.8 &  5.99 &  6.03 &  0.86 &  \ion{Mg}{ix} &  368.071 & 0.96 \\ 
 
 369.17 & 0.43 &  6.12 &  6.08 &  0.50 &  \ion{Fe}{xi} &  369.163 & 0.90 \\ 
 
 374.15 & 0.28 &  5.06 &  5.13 &  0.85 &  \ion{N}{iii} &  374.434 & 0.12 \\ 
                                 &   &  &  &  &  \ion{O}{iii} &  374.432 & 0.10 \\ 
                                 &   &  &  &  &  \ion{O}{iii} &  373.803 & 0.11 \\ 
                                 &   &  &  &  &  \ion{O}{iii} &  374.073 & 0.32 \\ 
 
 376.59 & 0.26 &  6.17 &  6.08 &  0.03 &  \ion{Fe}{xii} &  376.481 & 0.57 \\ 
                                 &   &  &  &  &  \ion{Mn}{ix} &  376.779 & 0.34 \\ 
 
 383.95 & 0.23 &  5.09 &  5.83 &  0.41 &  \ion{C}{iv} &  384.031 & 0.14 \\ 
                                 &   &  &  &  &  \ion{C}{iv} &  384.174 & 0.25 \\ 
                                 &   &  &  &  &  \ion{Al}{viii} &  383.778 & 0.40 \\ 
                                 &   &  &  &  &  \ion{Al}{viii} &  383.639 & 0.11 \\ 
 
 386.00 & 0.08 &  4.98 &  5.63 &  1.00 &  \ion{C}{iii} &  386.203 & 0.66 \\ 
                                 &   &  &  &  &  \ion{Ti}{xi} &  386.141 & 0.31 \\ 
 
 387.82 & 0.10 &  5.24 &  5.85 &  1.08 &  \ion{N}{iv} &  387.355 & 0.12 \\ 
                                 &   &  &  &  &  \ion{Al}{viii} &  387.952 & 0.49 \\ 
                                 &   &  &  &  &  \ion{Ne}{iv} &  388.225 & 0.12 \\ 
 
 392.44 & 0.11 &  6.03 &  6.03 &  0.86 &  \ion{Al}{ix} &  392.357 & 0.11 \\ 
                                 &   &  &  &  &  \ion{Al}{ix} &  392.403 & 0.83 \\

 399.89 & 0.15 &  5.61 &  5.66 &  0.76 &  \ion{Ne}{vi} &  399.841 & 0.97 \\ 
 
 400.76 & 0.28 &  5.62 &  5.71 &  1.27 &  \ion{Ne}{vi} &  401.146 & 0.62 \\ 
                                 &   &  &  &  &  \ion{Mg}{vi} &  400.663 & 0.30 \\ 
 
 401.85 & 0.66 &  5.61 &  5.66 &  0.84 &  \ion{Ne}{vi} &  401.941 & 0.97 \\ 
 
 403.17 & 0.23 &  5.63 &  5.72 &  1.28 &  \ion{Ne}{vi} &  403.260 & 0.43 \\ 
                                 &   &  &  &  &  \ion{Mg}{vi} &  403.308 & 0.55 \\ 
 
 411.15 & 0.23 &  5.87 &  5.94 &  0.53 &  \ion{Na}{viii} &  411.171 & 0.93 \\ 
 
 416.16 & 0.21 &  5.45 &  5.48 &  1.03 &  \ion{Ne}{v} &  416.210 & 0.96 \\ 
 
 419.60 & 0.20 &  5.08 &  5.77 &  0.28 &  \ion{C}{iv} &  419.525 & 0.19 \\ 
                                 &   &  &  &  &  \ion{C}{iv} &  419.714 & 0.38 \\ 
                                 &   &  &  &  &  \ion{Ca}{x} &  419.753 & 0.39 \\ 
 
 430.41 & 0.75 &  5.91 &  5.98 &  0.84 &  \ion{Mg}{viii} &  430.454 & 0.97 \\ 
 
 431.26 & 0.27 &  5.79 &  5.90 &  0.96 &  \ion{Mg}{vii} &  431.319 & 0.77 \\ 
                                 &   &  &  &  &  \ion{Mg}{vii} &  431.194 & 0.21 \\

 434.94 & 0.65 &  5.79 &  5.89 &  0.62 &  \ion{Mg}{vii} &  434.726 & 0.12 \\ 
                                 &   &  &  &  &  \ion{Mg}{vii} &  434.923 & 0.83 \\ 
 
 436.64 & 1.3 &  5.91 &  5.98 &  0.87 &  \ion{Mg}{viii} &  436.733 & 0.88 \\

 443.80 & 0.25 &  5.99 &  6.03 &  1.00 &  \ion{Mg}{ix} &  443.404 & 0.16 \\ 
                                 &   &  &  &  &  \ion{Mg}{ix} &  443.973 & 0.78 \\

 459.59 & 0.27 &  4.99 &  4.93 &  1.11 &  \ion{C}{iii} &  459.466 & 0.11 \\ 
                                 &   &  &  &  &  \ion{C}{iii} &  459.514 & 0.23 \\ 
                                 &   &  &  &  &  \ion{C}{iii} &  459.627 & 0.47 \\ 
 
 465.17 & 2.6 &  5.72 &  5.82 &  1.00 &  \ion{Ne}{vii} &  465.221 & 0.98 \\ 
 
 466.10 & 0.45 &  5.81 &  5.93 &  0.68 &  \ion{Ca}{ix} &  466.240 & 0.98 \\ 
 
 469.83 & 0.26 &  5.26 &  5.28 &  0.90 &  \ion{Ne}{iv} &  469.875 & 0.36 \\ 
                                 &   &  &  &  &  \ion{Ne}{iv} &  469.825 & 0.55 \\

 481.30 & 0.12 &  5.44 &  5.53 &  1.14 &  \ion{Ne}{v} &  481.374 & 0.38 \\ 
                                 &   &  &  &  &  \ion{Ne}{v} &  481.363 & 0.25 \\ 
                                 &   &  &  &  &  \ion{Ne}{v} &  481.291 & 0.32 \\ 
 
 482.96 & 0.20 &  5.44 &  5.48 &  1.17 &  \ion{Ne}{v} &  482.997 & 0.74 \\ 
                                 &   &  &  &  &  \ion{Ne}{v} &  482.985 & 0.24 \\ 
 
 489.48 & 0.15 &  5.06 &  5.15 &  1.18 &  \ion{Ne}{iii} &  489.495 & 0.79 \\ 
                                 &   &  &  &  &  \ion{Ne}{iii} &  489.629 & 0.15 \\ 
 
 499.41 & 0.34 &  6.29 &  6.20 &  0.95 &  \ion{Si}{xii} &  499.406 & 0.94 \\ 
 
 507.95 & 1.1 &  5.02 &  5.00 &  1.16 &  \ion{O}{iii} &  507.388 & 0.11 \\ 
                                 &   &  &  &  &  \ion{O}{iii} &  507.680 & 0.33 \\ 
                                 &   &  &  &  &  \ion{O}{iii} &  508.178 & 0.55 \\ 
 
 515.58 & 0.27 &  4.53 &  5.03 &  0.22 &  \ion{He}{i} &  515.618 & 0.88 \\ 
 
 520.68 & 0.18 &  6.29 &  6.21 &  0.88 &  \ion{Si}{xii} &  520.665 & 0.96 \\ 
 
 522.17 & 0.45 &  5.26 &  4.76 &  0.54 &  \ion{He}{i} &  522.214 & 0.85 \\ 
 
 525.79 & 0.66 &  5.03 &  5.00 &  0.90 &  \ion{O}{iii} &  525.794 & 0.99 \\ 
 
 537.01 & 1.4 &  4.52 &  4.54 &  0.43 &  \ion{He}{i} &  537.031 & 0.98 \\ 
 
 538.21 & 0.44 &  4.97 &  4.85 &  1.59 &  \ion{C}{iii} &  538.149 & 0.21 \\ 
                                 &   &  &  &  &  \ion{C}{iii} &  538.312 & 0.36 \\ 
                                 &   &  &  &  &  \ion{O}{ii} &  538.264 & 0.20 \\ 
                                 &   &  &  &  &  \ion{O}{ii} &  537.832 & 0.12 \\ 
 
 539.28 & 0.28 &  4.78 &  4.79 &  1.56 &  \ion{O}{ii} &  539.549 & 0.40 \\ 
                                 &   &  &  &  &  \ion{O}{ii} &  539.086 & 0.57 \\ 
 
 541.85 & 0.19 &  5.25 &  5.33 &  0.77 &  \ion{Ne}{iv} &  542.076 & 0.96 \\ 
 
 543.87 & 0.25 &  5.25 &  5.29 &  0.89 &  \ion{Ne}{iv} &  543.886 & 0.97 \\ 
 
 550.08 & 0.12 &  6.18 &  6.14 &  0.89 &  \ion{Al}{xi} &  550.031 & 0.94 \\ 
 
 553.35 & 0.99 &  5.22 &  5.24 &  0.85 &  \ion{O}{iv} &  553.329 & 0.97 \\ 
 
 554.39 & 5.5 &  5.22 &  5.23 &  1.06 &  \ion{O}{iv} &  554.076 & 0.27 \\ 
                                 &   &  &  &  &  \ion{O}{iv} &  554.514 & 0.70 \\ 
 
 555.27 & 0.77 &  5.22 &  5.20 &  1.19 &  \ion{O}{iv} &  555.264 & 0.90 \\ 
 
 557.60 & 0.26 &  5.89 &  6.03 &  1.30 &  \ion{Ca}{x} &  557.766 & 0.99 \\ 
 
 558.41 & 0.54 &  5.60 &  5.67 &  0.96 &  \ion{Ne}{vi} &  558.603 & 0.89 \\ 
 
 561.65 & 0.22 &  5.72 &  5.81 &  0.84 &  \ion{Ne}{vii} &  561.378 & 0.15 \\ 
                                 &   &  &  &  &  \ion{Ne}{vii} &  561.730 & 0.75 \\ 
 
 562.78 & 0.61 &  5.60 &  5.65 &  0.82 &  \ion{Ne}{vi} &  562.805 & 0.81 \\ 
                                 &   &  &  &  &  \ion{Ne}{vi} &  562.711 & 0.17 \\ 
 
 568.38 & 0.10 &  5.43 &  5.90 &  1.32 &  \ion{Al}{xi} &  568.120 & 0.43 \\ 
                                 &   &  &  &  &  \ion{Ne}{v} &  568.422 & 0.53 \\ 
 
 569.76 & 0.18 &  5.43 &  5.47 &  1.14 &  \ion{Ne}{v} &  569.837 & 0.75 \\ 
                                 &   &  &  &  &  \ion{Ne}{v} &  569.759 & 0.23 \\ 
 
 572.29 & 0.31 &  5.43 &  5.47 &  1.10 &  \ion{Ne}{v} &  572.336 & 0.84 \\ 
                                 &   &  &  &  &  \ion{Ne}{v} &  572.113 & 0.13 \\ 
 
 574.05 & 0.28 &  5.00 &  5.91 &  0.83 &  \ion{C}{iii} &  574.281 & 0.22 \\ 
                                 &   &  &  &  &  \ion{Ca}{x} &  574.008 & 0.73 \\ 
 
 580.93 & 0.16 &  6.22 &  6.03 &  0.92 &  \ion{Si}{xi} &  580.919 & 0.71 \\ 
                                 &   &  &  &  &  \ion{O}{ii} &  580.971 & 0.22 \\ 
 
 584.31 & 14.2 &  4.50 &  4.50 &  0.82 &  \ion{He}{i} &  584.335 & 0.98 \\ 
 
 585.67 & 0.13 &  5.55 &  5.57 &  1.32 &  \ion{Ar}{vii} &  585.748 & 0.79 \\ 
 
 599.59 & 1.5 &  5.02 &  4.99 &  1.07 &  \ion{O}{iii} &  599.590 & 0.99 \\ 
 
 608.39 & 0.70 &  5.21 &  5.25 &  1.12 &  \ion{O}{iv} &  608.397 & 0.97 \\ 
 
 609.79 & 5.0 &  5.21 &  5.93 &  0.87 &  \ion{Mg}{x} &  609.794 & 0.65 \\ 
                                 &   &  &  &  &  \ion{O}{iv} &  609.830 & 0.33 \\

 624.94 & 2.1 &  6.07 &  6.07 &  0.76 &  \ion{Mg}{x} &  624.968 & 0.92 \\ 
 
 629.73 & 13.9 &  5.39 &  5.38 &  1.04 &  \ion{O}{v} &  629.732 & 0.98 \\

 661.40 & 0.25 &  5.06 &  5.05 &  1.24 &  \ion{S}{iv} &  661.420 & 0.88 \\ 
 
 681.30 & 0.25 &  4.87 &  5.73 &  1.15 &  \ion{Na}{ix} &  681.719 & 0.46 \\ 
                                 &   &  &  &  &  \ion{S}{iii} &  681.488 & 0.11 \\ 
                                 &   &  &  &  &  \ion{S}{iii} &  680.973 & 0.26 \\ 
                                 &   &  &  &  &  \ion{S}{iii} &  680.924 & 0.14 \\ 
 
 685.77 & 1.2 &  4.97 &  4.91 &  1.30 &  \ion{N}{iii} &  685.515 & 0.24 \\ 
                                 &   &  &  &  &  \ion{N}{iii} &  686.336 & 0.12 \\ 
                                 &   &  &  &  &  \ion{N}{iii} &  685.817 & 0.62 \\

 702.25 & 0.32 &  4.99 &  4.97 &  1.64 &  \ion{O}{iii} &  702.337 & 0.96 \\ 
 
 702.84 & 1.5 &  4.99 &  4.93 &  1.08 &  \ion{O}{iii} &  702.838 & 0.29 \\ 
                                 &   &  &  &  &  \ion{O}{iii} &  702.896 & 0.24 \\ 
                                 &   &  &  &  &  \ion{O}{iii} &  702.900 & 0.39 \\ 
 
 703.85 & 2.1 &  4.99 &  4.96 &  1.21 &  \ion{O}{iii} &  703.851 & 0.25 \\ 
                                 &   &  &  &  &  \ion{O}{iii} &  703.854 & 0.74 \\ 
 
 706.05 & 0.52 &  5.99 &  6.02 &  0.78 &  \ion{Mg}{ix} &  706.060 & 0.95 \\

 718.52 & 0.63 &  4.77 &  4.68 &  3.05 &  \ion{O}{ii} &  718.504 & 0.55 \\ 
                                 &   &  &  &  &  \ion{O}{ii} &  718.566 & 0.36 \\

 759.40 & 0.34 &  5.39 &  5.38 &  0.70 &  \ion{O}{v} &  759.442 & 0.98 \\ 
 
 760.43 & 0.92 &  5.39 &  5.38 &  1.19 &  \ion{O}{v} &  760.227 & 0.16 \\ 
                                 &   &  &  &  &  \ion{O}{v} &  760.446 & 0.82 \\ 
 
 762.02 & 0.25 &  5.39 &  5.38 &  1.17 &  \ion{O}{v} &  762.004 & 0.98 \\ 
 
 765.11 & 3.7 &  5.18 &  5.16 &  0.92 &  \ion{N}{iv} &  765.152 & 0.98 \\ 
 
 770.41 & 3.6 &  5.80 &  5.98 &  1.10 &  \ion{Ne}{viii} &  770.428 & 0.99 \\ 
 
 780.31 & 2.1 &  5.80 &  5.96 &  1.01 &  \ion{Ne}{viii} &  780.385 & 0.95 \\ 
 
 786.49 & 1.5 &  5.22 &  5.21 &  0.98 &  \ion{S}{v} &  786.468 & 0.98 \\ 
 
 787.72 & 2.7 &  5.20 &  5.22 &  1.25 &  \ion{O}{iv} &  787.710 & 0.98 \\ 
 
 790.20 & 5.4 &  5.20 &  5.21 &  1.25 &  \ion{O}{iv} &  790.201 & 0.89 \\ 
 
 833.42 & 3.9 &  4.97 &  4.81 &  1.99 &  \ion{O}{iii} &  833.749 & 0.32 \\ 
                                 &   &  &  &  &  \ion{O}{iii} &  832.929 & 0.13 \\ 
                                 &   &  &  &  &  \ion{O}{ii} &  833.330 & 0.43 \\ 
 
 834.41 & 1.6 &  4.72 &  4.61 &  3.30 &  \ion{O}{ii} &  834.466 & 0.98 \\ 
 
 835.26 & 3.6 &  4.97 &  4.91 &  1.53 &  \ion{O}{iii} &  835.092 & 0.14 \\ 
                                 &   &  &  &  &  \ion{O}{iii} &  835.289 & 0.83 \\

 977.04 & 51.4 &  4.94 &  4.80 &  0.66 &  \ion{C}{iii} &  977.020 & 0.99 \\ 
 
 991.60 & 3.4 &  4.94 &  4.85 &  0.77 &  \ion{N}{iii} &  991.577 & 0.87 \\

1031.94 & 19.7 &  5.48 &  5.75 &  0.45 &  \ion{O}{vi} & 1031.912 & 0.98 \\ 
 
1036.54 & 3.1 &  4.65 &  4.55 &  2.70 &  \ion{C}{ii} & 1036.337 & 0.33 \\ 
                                 &   &  &  &  &  \ion{C}{ii} & 1037.018 & 0.66 \\ 
 
1037.58 & 11.1 &  5.48 &  5.76 &  0.39 &  \ion{O}{vi} & 1037.614 & 0.99 \\

\end{longtable}

\normalsize

\end{document}